\documentclass[aip,amsmath,amssymb,reprint]{revtex4-2} 	
\draft 

\usepackage{booktabs} 
\usepackage{bbold}		
\usepackage{graphicx}
\usepackage{dcolumn}
\usepackage{bm}
\usepackage[utf8]{inputenc}
\usepackage[T1]{fontenc}
\usepackage{mathptmx}
\usepackage{etoolbox}

\makeatletter
\def\@email#1#2{%
 \endgroup
 \patchcmd{\titleblock@produce}
  {\frontmatter@RRAPformat}
  {\frontmatter@RRAPformat{\produce@RRAP{*#1\href{mailto:#2}{#2}}}\frontmatter@RRAPformat}
  {}{}
}%
\makeatother
\begin{document}

\title{Isotope study of the nonlinear pressure shifts of $^{85}$Rb and $^{87}$Rb hyperfine resonances in Ar, Kr, and Xe buffer gases} 
\author{B. H. McGuyer} 
\email{bmcguyer@amazon.com.} 
\altaffiliation{Present address: Amazon.com, Inc., 410 Terry Ave. North, Seattle, WA 98109.} 
\affiliation{Department of Physics, Princeton University, Princeton, New Jersey 08544, USA}
\date{\today} 

\begin{abstract} 
Measurements of the 0--0 hyperfine resonant frequencies of ground-state $^{85}$Rb atoms show a nonlinear dependence on the pressure of the buffer gases Ar, Kr, and Xe. 
The nonlinearities are similar to those previously observed with $^{87}$Rb and $^{133}$Cs and presumed to come from 
alkali-metal--noble-gas van der Waals molecules. 
However, the shape of the nonlinearity observed for Xe 
conflicts  
with previous theory, and the nonlinearities for Ar and Kr 
disagree 
with the expected isotopic scaling of previous $^{87}$Rb results. 
Improving the modeling alleviates most of these discrepancies by treating rotation quantum mechanically and considering additional spin interactions in the molecules. 
Including the dipolar-hyperfine interaction allows simultaneous fitting of the linear and nonlinear shifts of both $^{85}$Rb and $^{87}$Rb in either Ar, Kr, or Xe buffer gases with a minimal set of shared, isotope-independent parameters. 
To the limit of experimental accuracy, the shifts in He and N$_2$ were linear with pressure. 
The results are of practical interest to vapor-cell atomic clocks and related devices. 
\end{abstract}
\maketitle

\tableofcontents

\begin{figure}[b!]
\centering
\includegraphics[width=\columnwidth]{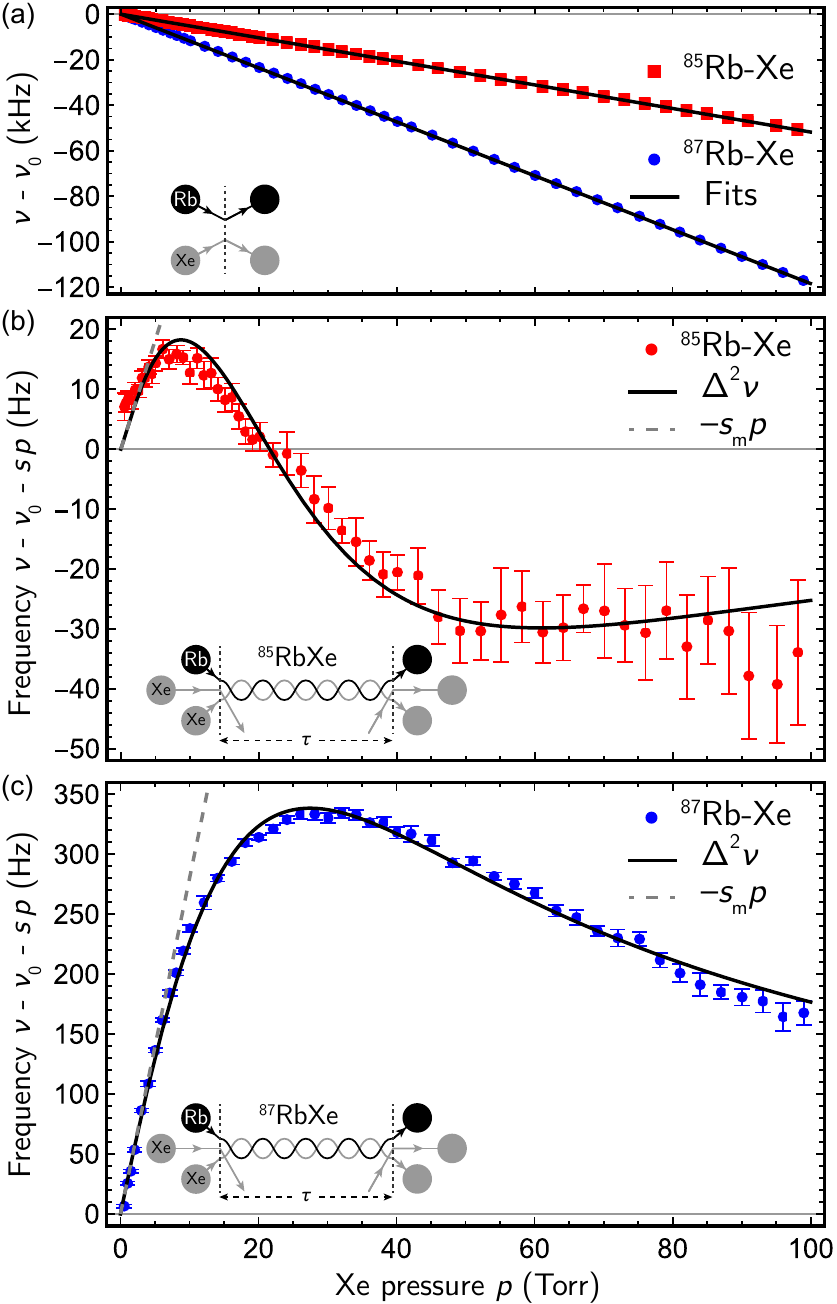}
\caption{ 	\label{fig1} 
Measured 0--0 resonant frequencies $\nu$ of $^{85}$Rb and $^{87}$Rb in Xe at 40.0$^\circ$ C and $B = 1$ G. 
(a) The pressure shifts $\nu - \nu_0$ from the free-atom frequencies $\nu_0$ are very nearly linear with pressure $p$. 
(b,c) Fitting and removing the linear, limiting shifts $s p$ at high pressures reveals nonlinear pressure shifts $\nu - \nu_0 - s p$. 
The solid curves are fitted nonlinear shifts $\Delta^2 \nu$ of (\ref{shiftFinal}) due to van der Waals molecules. 
Fitting used shared isotope-independent parameters, with fitting residuals given in Fig.~\ref{fig4}. 
The dashed lines correspond to the total linear shift, $s p$, minus the contribution from molecules, $s_m p$. 
The insets illustrate the dominant two- and three-body collision processes responsible for the pressure shift. 
}
\label{overview}
\end{figure}

\section{Introduction} 

The study of the interaction between alkali-metal atoms and chemically inert buffer gases has a long history because of its fundamental interest  
and its significance to many applications, 
from pressure measurement\cite{klos:2023} 
to laser guidestars.\cite{loreau:2011} 
In particular, this interaction is important to atomic devices like frequency standards (or clocks) and magnetometers that are based on the microwave resonance frequencies of dilute vapors of alkali-metal atoms in cells filled with buffer gasses, such as the Rb vapor-cell (or gas cell) clocks on board Global Positioning System satellites.\cite{vanier:1989,camparo:2007,kitching:2018,RbPrimer} 
In these devices, collisions with the gas shift the resonant frequencies $\nu$ away from their free-atom values $\nu_0$ by an amount, $\nu - \nu_0$, that depends on the gas composition, pressure, and temperature. 
Known as the ``pressure'' shift,\cite{ishikawa:2023,ishikawa:2022,camparo:2007jcp,oreto:2004,ha:2021} 
these shifts are very nearly linear with the buffer-gas pressure $p$ at typical partial-vacuum cell pressures. 
Fig.~\ref{fig1}(a) provides an example for the ``0--0'' (or clock) frequencies of $^{85}$Rb and $^{87}$Rb in Xe gas. 
Two-body (or binary) collisions, such as Rb + Xe $\longrightarrow$ Rb + Xe, that briefly but repetitively alter the hyperfine coupling in each alkali-metal atom produce the majority of this pressure shift. 
The pressure shift and its associated broadening impact the accuracy and performance of vapor-cell clocks. 

Much of this study has focused on the interaction between alkali-metal atoms and the noble gases.\cite{medvedev:2018,patil:1991,pascale:1974,buckpauli:1968} 
In part, this is because these collision pairs readily form weakly bound van der Waals molecules,\cite{tariq:2013,brahms:2010,dhiflaoui:2011,kerkines:2002,partridge:2001,dehmer:1972,ury:1972} 
which have long been known to be important to atomic devices.\cite{happer:1972,bouchiat:1972, bouchiat:1975} 
Three-body ``sticking'' collisions (or teratomic recombination), \cite{mirahmadi:2021a,mirahmadi:2021b,haze:2022,reynolds:2020,wolf:2019,wolf:2017,scheps:1976} 
such as Rb + Xe + Xe $\longrightarrow$ RbXe + Xe, produce the majority of these short-lived molecules, which remain until later dissociating, primarily in two-body atom-molecule collisions (or time-reversed teratomic recombination).  
Though molecule-forming collisions are relatively rare at typical pressures, their effects can still be significant because the resulting molecules persist drastically longer than the duration of binary collisions. 
For example, the spin-rotation interaction between molecular rotation and the alkali-metal electronic spin is often a significant source of spin relaxation in vapor cells.\cite{wu:1985,happer:book}  

Returning to the pressure shift, precise measurements discovered a small nonlinear dependence on the pressure for $^{87}$Rb and $^{133}$Cs in the heavy noble gases Ar, Kr, and Xe, as Fig.~\ref{fig1} shows.\cite{gong:2008,mcguyer:2011pra,mcguyer:thesis}  
Modeling explained these observations by considering spin interactions in alkali-metal--noble-gas molecules formed by sticking collisions, and revealed that the observed nonlinearity primarily came from the gradual turning on of a linear shift from the molecules with increasing pressure. 
To date, no comprehensive tests of these models have been reported, in particular, using data from other alkali-metal atoms. 
Measuring these nonlinear pressure shifts provides one of the few avenue to probe such van der Waals molecules, 
which are relevant to clocks,\cite{gozzelino:2021} 
spin-exchange optical pumping,\cite{walker:1997,gentile:2017} 
cold molecule production,\cite{cantano:2015,quiros:2017,tariq:2013,brahms:2010} 
quantum memories,\cite{chrapkiewicz:2014,wasilewski:2014} 
and other areas. 

To advance the understanding of spin interactions in alkali-metal--buffer-gas van der Waals molecules, this Article reports an isotope comparison of the 0--0 pressure shifts of $^{85}$Rb and $^{87}$Rb, which uses measurements with $^{85}$Rb in He, N$_2$, Ar, Kr, natural Xe, and spinless $^{136}$Xe buffer gases together with previous $^{87}$Rb data. 
Nonlinear pressure shifts are present for both isotopes of Rb in Ar, Kr, and Xe, but their variation with isotope disagrees with previous theory. 
The disagreement is most striking for $^{85}$Rb in Xe, whose nonlinearity has a shape shown in Fig.~\ref{fig1} that is incompatible with the shapes allowed by previous modeling, which generally resemble the shape shown in Fig.~\ref{fig1}(c). 
To resolve these discrepancies, this Article presents an improved model that includes additional spin interactions in the molecules and treats rotation quantum mechanically, instead of semi-classically as in previous work. 
As shown in Fig.~\ref{fig1} for Xe, this model is able to simultaneously fit the data for both Rb isotopes in pure Ar, Kr, or Xe gas, including all linear and nonlinear shifts, by incorporating the dipolar-hyperfine interaction in addition to the hyperfine-shift and spin-rotation interactions considered previously. 

This Article is organized as follows: 
Section II provides a theoretical model for the linear and nonlinear pressure shifts of all hyperfine transitions in an alkali-metal atom from $X$ $^2{\Sigma}^+$ van der Waals molecules, reproducing all previous models except for the modifications of Camparo\cite{camparo:TOR,camparo:2022} discussed in Section V. 
The derivation includes the hyperfine-shift and spin-rotation interactions of previous work, but predicts an important correction to the spin-rotation contribution. 
Additionally, it includes the Zeeman interaction and the dipolar- and quadrupolar-hyperfine interactions. 
Section III describes the experimental measurement of 
the pressure shift data and the mitigation of systematic effects. 
Section IV reports analysis of $^{85}$Rb and $^{87}$Rb data in He, N$_2$, Ar, Kr, and Xe gases, as well as $^{85}$Rb data in Xe and $^{136}$Xe. 
The data for He and N$_2$ were linear to within experimental error, and provide a check of systematic errors. 
The model of Section II successfully captures the unusual shape of the nonlinearity of $^{85}$Rb in Xe, 
and successfully fits the nonlinear as well as the linear shifts of both Rb isotopes in each nonlinear gas. 
Section V discusses these results as well as a recent analysis of the Xe data by Camparo.\cite{camparo:TOR,camparo:2022} 
Section VI concludes the Article, and the 
Supplementary Material provide additional details, figures, and fitting parameters. 

\section{Theoretical Molecular Pressure Shift} 


Between collisions, a free alkali-metal atom evolves according to a ground-state spin Hamiltonian 
\begin{align}	\label{H0}
H_0 = A \, {\bf I} \cdot {\bf S} - \bm{\mu} \cdot {\bf B}, 
\end{align}
where the first term is a Fermi-contact hyperfine interaction coupling the nuclear spin ${\bf I}$ and electronic spin ${\bf S}$ with a magnetic-dipole coupling coefficient $A$. 
The second term is a Zeeman interaction of the total magnetic-dipole moment 
$\bm{\mu} = - g_S \mu_B {\bf S} +g_I \mu_N {\bf I}$ 
with an externally applied magnetic field ${\bf B}$ of amplitude $B = |{\bf B}|$. 
Here, $g_S$ is the electronic $g$ factor, $\mu_B$ is the Bohr magneton, 
$g_I$ is the nuclear $g$ factor, and $\mu_N$ is the nuclear magneton. 
Let us assume that any external field is static and oriented along the lab-frame Cartesian unit vector ${\bf z}$, such that ${\bf B} = B {\bf z}$. 

For vapor-cell clocks, the field $B$ is typically weak enough that the dominant interaction in $H_0$ is the hyperfine coupling. 
The ground-state energy eigenstates are very nearly the eigenstates $| F \, m \rangle$ of the total spin angular momentum ${\bf F} = {\bf I} + {\bf S}$ with quantum number $F$ and azimuthal quantum number $m$ along ${\bf z}$. 
The hyperfine coupling splits the sublevels into upper and lower hyperfine manifolds with total angular momentum  
\begin{align}
F = 
\begin{cases}
a = I +1/2, \\ 
b = I -1/2, 
\end{cases} 
\end{align} 
respectively. 
The nuclear spin $I$ and electronic spin $S = 1/2$ are good quantum numbers for these sublevels. 

Vapor-cell clocks and related devices measure transitions between these hyperfine manifolds. 
To proceed, let us consider an ``$\alpha$--$\beta$'' hyperfine transition between the free-atom eigenstates 
\begin{align}	\label{aabb}
|\alpha \rangle = |a \, \alpha \rangle 
\;
\text{ and } 
\;
|\beta \rangle = |b \, \beta \rangle. 
\end{align} 
Clocks conventionally use the $0$--$0$ transition ($\alpha = \beta = 0$) because it has no first-order dependence on the field $B$. 
Ideally, without collisions, the measured resonant frequency would be the Bohr angular frequency for the transition, 
\begin{align} 
\omega_{\alpha \beta} = ({E_{a \alpha} - E_{b \beta}})/{\hbar}, 
\end{align} 
where $\hbar$ is the reduced Planck constant and the energies $E_{F m} = \langle F \, m | H_0 | F \, m \rangle$. 
These frequencies are approximately  
\begin{align} 	\label{v00}
\omega_{\alpha \beta} 
	= 2 \pi \, \nu_{\alpha\beta} 
	\approx \frac{A[I]}{2\hbar} + \frac{g_S \mu_B B}{[I]\hbar} (\alpha + \beta), 
\end{align} 
to first order in the field $B$ and ignoring the small Zeeman interaction with the nuclear spin. 
Here and subsequently, the shorthand of brackets about a single quantum number,
\begin{align}
[J] = 2 J + 1,
\end{align}  
denotes its spin multiplicity. 
Table~\ref{tab1} provides values for $\nu_{00}$ and other atomic parameters of interest to fitting functions.

\begin{table}[t]
\caption{
Atomic parameters for fitting functions. 
The unperturbed, ground hyperfine frequencies $\nu_{00}$ are rounded to the nearest Hz. 
The values of $g_I$ are from nuclear magnetic-dipole moments $\mu_I = g_I I \mu _N$ in Ref.~\onlinecite{stone}. 
The values of $Q$ are inferred from molecules, and expected to vary slightly with the particular molecule.\cite{kello:1999,pyyko} 
\label{tab1} 
}
\begin{ruledtabular}
\begin{tabular}{ l l l l l } 
Atom 		& $I$ 	& $\nu_{00}$ (Hz) 
	& $g_I$ [Ref.~\onlinecite{stone}] 		
		& $Q$ (Barns) [Ref.~\onlinecite{kello:1999}] \\ 
\midrule 
$^{85}$Rb 	& 5/2	& 3,035,732,439 
	& +0.54119		
		& $+0.276 \pm 0.002$ 	\\ 
$^{87}$Rb 	& 3/2	& 6,834,682,611 
	& +1.83421		
		& $+0.134 \pm 0.002$ 
\end{tabular}	
\end{ruledtabular}
\end{table}

\subsection{Pressure shifts from sticking collisions} 

We can calculate the pressure shift from molecule-forming collisions using a density-matrix approach as follows.\cite{mcguyer:thesis}  
The density operator $\rho$ for ground-state alkali-metal atoms evolves according to 
\begin{align}
\frac{\partial \rho}{\partial t} 
	= - \frac{i}{\hbar} [ H_0, \rho ] 
	+ \left\langle \frac{1}{T}  \left( \hat{S} \rho \hat{S}^\dagger - \rho \right) \right\rangle, 
\end{align}
which for simplicity ignores the effects of optical and microwave interactions as well as binary and other non-sticking collisions. 
The first term on the right models free-atom evolution. 
The second term models sticking collisions with angle brackets denoting an ensemble average over molecules and their formation rates $1/T$. 
$\hat{S}$ is a scattering-matrix (S-matrix) operator that captures bound-atom evolution by converting the wave function of a free atom into that of a free atom just exiting a collisional perturbation, following Ch.~10 of Ref.~\onlinecite{happer:book}. 
For a particular molecule with lifetime $\tau$, this S-matrix is 
\begin{align} 
\hat{S} = \exp({-i {H_1} \tau/\hbar}) \exp({i H_0 \tau/\hbar}), 
\end{align} 
where ${H_1}$ is the spin Hamiltonian for the bound-atom evolution. 

The bound-atom spin Hamiltonian ${H_1}$ includes additional interactions that only contribute during the time spent in the molecule. 
For Rb, the most significant interactions are expected to be 
the hyperfine-shift (hfs), 
electronic spin-rotation (sr), 
dipolar-hyperfine (dh), and 
quadrupolar-hyperfine (qh) interactions, 
\begin{align}
{H_1} \approx H_0 + V_\text{hfs} + V_\text{sr} + V_\text{dh} + V_\text{qh}, 
\end{align}
which are each addressed below.\cite{walker:1997} 
Other interactions are assumed negligible, such as nuclear spin-rotation and octupole-hyperfine interactions. 
Interactions with spins in the bound partner like spin exchange are present, but as discussed below, measurements with natural Xe and spinless $^{136}$Xe agree, 
suggesting they are negligible for typical buffer gases. This may no longer apply if, for example, the buffer gas is spin polarized. 

Ideally, the measured $\alpha$--$\beta$ frequency is equal to the precession frequency of the coherence $\langle \alpha | \rho | \beta \rangle$. 
In the secular approximation, molecules from sticking collisions produce a pressure shift   
\begin{align}	\label{shift1}
\Delta \nu 
	= - \text{Im} \left\langle \frac{\langle \alpha | \hat{S} | \alpha \rangle \langle \beta | \hat{S}^\dagger | \beta \rangle }{ 2 \pi T }\right\rangle 
\end{align}
of the temporal frequency of this coherence as well as a corresponding damping.\cite{mcguyer:thesis}  
As before, angle brackets denote averaging over molecules. 

For a given rovibrational state, the lifetime is expected to follow an exponential distribution with mean $\tau$. 
We can average over molecular lifetimes by integrating
$\int_0^\infty e^{-t/\tau} \Delta \nu \, dt/\tau$ 
after introducing bound-atom spin eigenstates $|\overline{\mu}\rangle$ of ${H_1}$ with energies $E_{\overline{\mu}} = \langle \overline{\mu} | {H_1} | \overline{\mu} \rangle$. 
Averaging then reduces the shift (\ref{shift1}) to a sum over bound-atom-state indices $\overline{\mu}$ and $\overline{\nu}$, 
\begin{align} \label{shift2}
\Delta \nu 
	= \sum_{\overline{\mu}, \overline{\nu}} \left\langle \frac{ \left| \langle \alpha | \overline{\mu} \rangle \langle \overline{\nu} | \beta \rangle \right|^2 (\omega_{\overline{\mu} \, \overline{\nu}} - \omega_{\alpha \beta} ) \tau }{ 2 \pi T \, [1 + (\omega_{\overline{\mu} \, \overline{\nu}} - \omega_{\alpha \beta} )^2 \tau^2]} \right\rangle, 
\end{align} 
where the bound-atom Bohr frequencies are 
\begin{align} 	\label{bohr2}
\omega_{\overline{\mu} \, \overline{\nu}} = ({E_{\overline{\mu}} - E_{\overline{\nu}}})/{\hbar}.  
\end{align} 
Here, and subsequently in the next few sections, a bar is added to help distinguish bound-atom quantities and states from those of free atoms. 
This is because, as we will see below, the indices $\overline{\mu}$ and $\overline{\nu}$ are effectively rotated azimuthal quantum numbers $\overline{m}$, similar to $\alpha$ and $\beta$ for free atoms. As a result, both $\overline{\mu}$ and $\alpha \in [-a, a]$, and both $\overline{\nu}$ and $\beta \in [-b, b]$. However, $|\overline{\mu}\rangle$ and $|\overline{\nu}\rangle$ do not follow (\ref{aabb}), unlike $|{\alpha}\rangle$ and $|{\beta}\rangle$.  

Eq.~(\ref{shift2}) is the pressure shift from molecules. 
Its rough dependence on buffer-gas pressure $p$ follows from noting that the three-body formation rate $1/T \propto p^2$ and the collision-limited lifetime $\tau \propto 1/p$. 
At low pressures there is no shift, 
$\lim_{p \rightarrow 0} \Delta \nu \propto p^3$, 
but at high pressures there is a linear shift, 
$\lim_{p \rightarrow \infty} \Delta \nu \propto p$. 
Thus, the rough shape of $\Delta \nu$ with pressure is a gradual turning on of a linear shift, with a detailed shape in between that is sensitive to molecular parameters. 
One interpretation of the shape of $\Delta \nu$ versus the inverse lifetime $1/\tau \propto p$ is as an interference pattern formed by all of the ways that an $\alpha$--$\beta$ coherence connects through a molecule via discrete $\overline{\mu}$--$\overline{\nu}$ pathways during the lifetime $\tau$. 

In measurements, the dominant linear shift is typically that from binary collisions instead of sticking collisions.\footnote{In theory, a mixture of buffer gases could be adjusted to null the binary shift,\cite{mcguyer:thesis} leaving only a molecular shift}  
Therefore, it is convenient to artificially separate the molecular shift into linear and nonlinear parts, 
\begin{align} 	\label{shift3}
\Delta \nu = s_\text{m} \, p + \Delta^2 \nu, 
\end{align} 
to highlight the nonlinearity that is due to sticking collisions. 
Here, the infinite-pressure linear molecular slope is 
\begin{align} \label{sm1}
s_\text{m} 
	&= \lim_{p \longrightarrow \infty} \Delta \nu / p \nonumber \\ 
	&= \sum_{\overline{\mu}, \overline{\nu}} \left\langle \frac{ \left| \langle \alpha | \overline{\mu} \rangle \langle \overline{\nu} | \beta \rangle \right|^2 (\omega_{\overline{\mu} \, \overline{\nu}} - \omega_{\alpha \beta} ) \tau}{2 \pi T p} \right\rangle, 
\end{align} 
and the remaining nonlinear shift from molecules is 
\begin{align} \label{D2nu1}
\Delta^2 \nu 
	&= \Delta \nu - s_m \, p \nonumber \\ 
	&= - \sum_{\overline{\mu}, \overline{\nu}} \left\langle  \frac{ \left| \langle \alpha | \overline{\mu} \rangle \langle \overline{\nu} | \beta \rangle \right|^2 (\omega_{\overline{\mu} \, \overline{\nu}} - \omega_{\alpha \beta} )^3 \tau^3 }{ 2 \pi T \, [1 + (\omega_{\overline{\mu} \, \overline{\nu}} - \omega_{\alpha \beta} )^2 \tau^2 ]} \right\rangle. 
\end{align} 
By construction, the nonlinear shift is zero at infinite pressure, 
$\lim_{p \rightarrow \infty} \Delta^2 \nu \rightarrow 0$, 
and as a result, is linear at low pressure, 
$\lim_{p \rightarrow 0} \Delta^2 \nu = - s_\text{m} p$, as highlighted in Fig.~\ref{fig1} and later figures.  
To avoid confusion, note that $\Delta^2 \nu$ is not the complete pressure shift from molecules. 
The true molecular pressure shift is the $\Delta \nu$ of (\ref{shift2}). 
We will focus on the molecular pressure shift $\Delta \nu$ below and return to the linear, limiting molecular slope $s_\text{m}$ of (\ref{sm1}) and nonlinear shift $\Delta^2 \nu $ of (\ref{D2nu1}) to construct fitting functions. 

To calculate these shifts, what remains is to determine the bound-atom eigenstates $| \overline{\mu} \rangle$ and their energies $E_{\overline{\mu}}$ and to average over rovibrational states, their formation rates, and their directions of rotation.

\subsection{Connecting bound-atom and molecular-spin eigenstates} 
To determine the bound-atom eigenstates $|\overline{\mu}\rangle$, we must choose how to treat molecular rotation. 
Previous work\cite{gong:2008,mcguyer:2011pra,mcguyer:thesis,camparo:2022,camparo:TOR} 
approximated rotation semi-classically in the interactions and thus separately from quantum spins. 
However, the derivation of (\ref{shift2}) with an S-matrix requires the bound-atom $|\overline{\mu}\rangle$ to be energy eigenstates. 
While bound, the true energy eigenstates are the molecular spin eigenstates formed by the coupling of alkali-metal atomic spins with molecular rotation. 
Therefore, we will include rotation quantum mechanically in the molecular spin states, connect them with the bound-atom spin states, and then take a classical large-rotation limit. 
The energies for the interactions considered agree to leading order for both approaches, though their interpretation with respect to quantization is different.

The molecules of interest are loosely bound heteronuclear diatomic molecules in their electronic ground states, 
composed of a $^1S_0$ noble-gas atom and a $^2S_{1/2}$ alkali-metal atom, with  molecular term symbol $X$ $^2{\Sigma}^+$. 
Their total electronic spin angular momentum ${\bf S}$ is solely due to the alkali-metal atom, so the quantum number $S=1/2$. 
Their total orbital angular momentum ${\bf L}=0$, 
with axial component $\Lambda = 0$, 
so these molecules follow Hund's case (b$_{\beta S}$) with their electronic spin not strongly coupled to the internuclear axis.\cite{brown:book}
The total rotational angular momentum of the nuclei is ${\bf N}$ with quantum number $N$. 
As explained above, we will ignore the nuclear spin from the noble-gas atom. 

The dominant spin interaction is expected to be the Fermi-contact hyperfine coupling between ${\bf I}$ and ${\bf S}$, just as for the free alkali-metal atoms, so the resultant ${\bf F} = {\bf I} + {\bf S}$ is still a good quantum number. 
The total angular momentum for the molecule is then the resultant ${\bf G} = {\bf F} + {\bf N}$. 
Note that here, and subsequently, the definitions of ${\bf F}$ and ${\bf G}$ are intentionally swapped compared to those in Brown and Carrington,\cite{brown:book} which is frequently referenced below, so that ${\bf F}$ has the same definition for ground-state atoms as for molecules. 

As we will see below, the Zeeman interaction with ${\bf B}$ in (\ref{H0}) sets the quantization axis for the molecules, just as for free atoms. 
The remaining spin-rotation, dipolar-hyperfine, and quadrupolar-hyperfine interactions do not depend on the azimuthal quantum number $m_G$, just as the spin-orbit, dipolar-hyperfine, and quadrupolar-hyperfine interactions in excited alkali-metal atomic spin states do not depend on the total $m$. 
Thus, for up to moderate applied magnetic fields the molecular spin eigenstates are very nearly the eigenstates 
\begin{align}
| I S F; F N G; G \, m_G \rangle 
\end{align} 
of the total spin angular momentum ${\bf G}$ with quantum number $G$ and azimuthal quantum number $m_G$ along ${\bf z}$, 
in the notation of Brown and Carrington.\cite{brown:book}  
As will be shown below, $G$ plays the role of $\overline{m}$ in a rotated, bound-atom spin state, connecting this with previous work.\cite{mcguyer:thesis}  
For convenience, let us use the shorthand 
\begin{align}
| F N G \, g \rangle = | I S F; F N G; G \, g \rangle 
\end{align}
with $g = m_G$, 
when the values of $I$, $S$, and $N$ are understood. 
Note that for sufficiently strong applied magnetic fields, there will be mixing between $G$ states, just as between $F$ states for free atoms. 

To use these molecular eigenstates, first replace the bound-atom eigenstates in (\ref{shift2}) with the substitutions 
\begin{align} 		\label{sub1} 
|\overline{\mu}\rangle 
	&\longrightarrow |F=a, N, G, g \rangle \\ 
|\overline{\nu}\rangle 
	&\longrightarrow |F'=b, N, G', g' \rangle \label{subNu} \\ 
\sum_{\overline{\mu},\overline{\nu}} 
	&\longrightarrow \sum_{G, g, G', g'}, 
\end{align} 
where (\ref{aabb}) set $F$ and $F'$. 
Then, replace the bound-atom Bohr frequencies (\ref{bohr2}) and energies with the substitutions 
\begin{align} 		
\omega_{\overline{\mu} \, \overline{\nu}}  
	&\longrightarrow \omega_{G g; G' g'} = (E_{|a N G g \rangle} - E_{|b N G' g' \rangle})/{\hbar} \\ 
E_{\overline{\mu}} 	
	&\longrightarrow E_{|a N G g \rangle} =  \langle a N G \, g | {H_1} | a N G \, g \rangle \\ 
E_{\overline{\nu}} 
	&\longrightarrow E_{|b N G' g' \rangle} =  \langle b N G' \, g' | {H_1} | b N G' \, g' \rangle. 
\end{align} 
Finally, the free-atom spin eigenstates (\ref{aabb}) have to be modified to include a tensor product, 
\begin{align} 
|\alpha\rangle 	
	&\longrightarrow | a \, \alpha \rangle \otimes |\psi_N \rangle \\ 
|\beta\rangle 
	&\longrightarrow | b \, \beta \rangle \otimes |\psi_N \rangle, \label{sub25}
\end{align} 
with a rotational wave function $|\psi_N \rangle$ 
that will be used to average over the direction of rotation. 

\subsection{Hyperfine propensity rule} \label{hyperfinepropensityrule}

Before we continue, note that the above approximations assume an adiabatic propensity rule that conserves the alkali-metal hyperfine spin state in the three-body recombination processes of both molecule association and dissociation. 
This might be surprising because the recombination processes are necessarily nonadiabatic. 
However, these processes are driven by mechanical forces and at interparticle distances where the interaction is likely independent of spin, in particular, because the weakly bound van der Waals molecules are held together by long-range dispersion forces, not short-range bonds. 
Recent state-to-state recombination experiments with ultracold Rb support choosing this propensity rule.\cite{haze:2022} 

\begin{table}[t]
\caption{
Molecular parameters for reference from Bouchiat {\it et al.}\cite{bouchiat:1972, bouchiat:1975} 
The values given as ranges without uncertainties are theoretical estimates. 
$B_1 = \langle \gamma N \rangle / (g_S \mu_B)$ is an experimental, effective spin-rotation magnetic field. 
The bottom row estimates the magnitude of the spin-rotation parameter 
(\ref{PsiPsr}) 
as $|\psi p|_\text{sr} = g_S \mu_B B_1 \langle \tau p \rangle/\hbar$. 
\label{tab2}
} 
\begin{ruledtabular}
\begin{tabular}{ l l l l l } 
\multicolumn{2}{l}{Parameter} 
		& RbAr [Ref.~\onlinecite{bouchiat:1975}]		
		& RbKr [Ref.~\onlinecite{bouchiat:1972}]		
		& RbXe [Ref.~\onlinecite{bouchiat:1972}] \\ 
\midrule 
$\left\langle T p^2\right\rangle$ & (ms Torr$^2$)
		& 16.1 $\pm$ 1.3 	
		& 10.6 $\pm$ 0.5	
		& 4.29 $\pm$ 0.23 \\ 
$\langle \tau p \rangle$ & (ns Torr)
		& 48.5 $\pm$ 1.9	
		& 56.9 $\pm$ 1.7	
		& 34--61 \\ 
$\langle N \rangle$ & 
		& 30.5--33.3 
		& 41.5 
		& 63.2--76.7 \\ 
$B_1$ & (Gauss) 	
		& 1.19 $\pm$ 0.05 
		& 9.59 $\pm$ 0.28 
		& 38.1 $\pm$ 1.6 \\
$|\psi p|_\text{sr}$ & (rad Torr) 
		& 1.016 $\pm$ 0.040 
		& 9.61 $\pm$ 0.40 
		& 31.9 $\pm$ 9.2 
\end{tabular}	
\end{ruledtabular}
\end{table}

\subsection{Large-$N$ approximation} 
To simplify the matrix elements in (\ref{shift2}), we will take a classical limit of $N \gg 1$, which is justified for RbAr, RbKr, and RbXe as shown in Table~\ref{tab2}.\cite{bouchiat:1972,bouchiat:1975} 
The molecular spin eigenstates decompose into atomic and rotational parts as 
\begin{align} 	\label{expand}
|F N G \, g \rangle = \sum_{m, m_N} C^{G \, g}_{F m, N m_N} | F \, m \rangle \otimes | N \, m_N \rangle. 
\end{align} 
For large $N$ (and thus large $G$), 
the quantum number $F \ll G$ and $ N$, and the Clebsch-Gordon coefficient is approximately [c.f. Eq.~8.9.1(1) of Ref.~\onlinecite{varshalovich}]  
\begin{align}	\label{CGlargeN}
\left| C^{G \, g}_{F m, N m_N} \right| 
	\approx \delta_{g, m + m_N}  \left| d^{(F)}_{m, G-N} \left(\theta_{G g} \right) \right|
\end{align}
where the angle between ${\bf G}$ and the lab ${\bf z}$ axis is 
\begin{align}	\label{thetaGg}
\theta_{G g} = \arccos( 2 g / [G] ). 
\end{align}
The Wigner ``little'' $d$-function\cite{varshalovich} is 
\begin{align}	\label{mAha}
d^{(F)}_{m, \overline{m}}(\theta) = \langle F \, m | \exp(- i \theta F_y) | F \, \overline{m}\rangle
\end{align}
where the the azimuthal quantum number  
\begin{align}
\overline{m} = G - N \in [-F,F], 
\end{align}
is that of an effective bound-atom state $|F \, \overline{m}\rangle$. 
The approximation (\ref{CGlargeN}) is surprisingly adequate for values of $N$ that are far from asymptotic, such as those in fitting results to come. 

Using (\ref{expand}) and (\ref{CGlargeN}) with (\ref{sub1}) and (\ref{subNu}), 
the matrix elements in (\ref{shift2}) simplify to 
\begin{align} 	\label{sub31}
|\langle \alpha | \overline{\mu}\rangle|^2 
	&\longrightarrow  
		\left|d^{(a)}_{\alpha, \overline{\mu}}(\theta_{G g})  \langle \psi_N | N, g - \alpha \rangle \right|^2 \\ 
|\langle \overline{\nu} | \beta \rangle|^2 
	&\longrightarrow 
		\left|d^{(b)}_{\beta, \overline{\nu}}(\theta_{G' g'})  \langle \psi_N | N, g' - \beta \rangle \right|^2 \label{sub32}
\end{align}
for large $N$. 
Here, the indices 
$\overline{\mu} = G - N \in [-a,a]$ and 
$\overline{\nu} = G' - N \in [-b,b]$, or equivalently $G$ and $G'$,  
play the role of $\overline{m}$ for rotated, bound-atom spin eigenstates, as in previous work.\cite{mcguyer:thesis,mcguyer:2011pra,camparo:2022,camparo:TOR}

\subsection{Averaging over the direction of rotation} 	\label{AverageOverDirection} 
To average over the direction of rotation (see Supplementary Material), replace the rotational wave function with the substitution 
\begin{align} 	\label{subPsiN}
|\psi_N\rangle \longrightarrow | N \, n \rangle
\end{align} 
and uniformly average over $n$. 
Note that for large $N$, $G \approx G' \approx N$ and $g \approx g' \approx n$ in $\Delta\nu$. 
As a result, the angles $\theta_{G g} \approx \theta_{G' g'}$ and are approximately continuous. 
After the substitution (\ref{subPsiN}), this average is then approximately
\begin{align} 
\bigg\langle \Delta\nu \bigg\rangle_{\psi_N} 
	= \frac{1}{[N]} \sum_{n = -N}^N \Delta\nu 
	\approx \frac{1}{2} \int_{-1}^1 \Delta\nu \; d\cos(\theta) 
\end{align} 
with a shared angle 
\begin{align} \label{sharedangle}
\theta \approx \theta_{G g} \approx \theta_{G' g'}.
\end{align}  

Together with the substitutions (\ref{sub1})--(\ref{sub25}), (\ref{sub31}), and (\ref{sub32}), this makes the 
nonlinear pressure shift (\ref{shift2}) 
\begin{align} 	\label{shift3}
\Delta \nu 
	&= \sum_{\overline{\mu}, \overline{\nu}} \int_{-1}^1  \left\langle \frac{ 
	f_{\overline{\mu} \, \overline{\nu}}^{\alpha \beta} (\theta) 
	(\omega_{\overline{\mu} \, \overline{\nu}}(\theta) - \omega_{\alpha \beta} ) \tau }{4 \pi T [ 1 + ( \omega_{\overline{\mu} \, \overline{\nu}}(\theta) - \omega_{\alpha \beta} )^2 \tau^2 ]} \right\rangle d\cos(\theta),  
\end{align} 
where the angular weight functions 
\begin{align} 
f_{\overline{\mu} \, \overline{\nu}}^{\alpha \beta} (\theta) 
	&= \left|d^{(a)}_{\alpha, \overline{\mu}}(\theta) \, d^{(b)}_{\beta, \overline{\nu}}(\theta) \right|^2
\end{align}
generalize 
those in Eq.~(4.94) of Ref.~\onlinecite{mcguyer:thesis}. 
The frequencies  
 \begin{align} 
\omega_{\overline{\mu} \, \overline{\nu}}(\theta) 
	= [&E_{|a, N, G = \overline{\mu}+N, g = n + \alpha \rangle}(\theta) \nonumber \\ 
	- &E_{|b, N, G' = \overline{\nu}+N, g' = n + \beta \rangle} (\theta) ]/{\hbar} 
\end{align} 
depend on the shared angle $\theta$ of (\ref{sharedangle}) and (\ref{thetaGg}) if the energies depend on 
$n \approx \cos(\theta) [N]/2$ via $g$ or $g'$, 
which occurs only for the Zeeman interaction as shown below.

\subsection{Single-rovibrational-state approximation} \label{singlestateapproximation}

The remaining average to complete is that over all rotational and vibrational states allowed by the alkali-metal--noble-gas interaction potential $V(R)$.\cite{medvedev:2018,patil:1991,pascale:1974,buckpauli:1968} 
This average superposes the differing shifts from each rovibrational state. 
Unfortunately, there is not enough information currently available about the spin interactions of interest for this to be tractable or trustworthy.  
Instead, to proceed we will approximate this average using a single rovibrational state with effective parameters, following all previous work. 
That is, we will assume the average is performed over the parameters, keeping the functional form of the shift unchanged. 

In this approximation, the $\Delta \nu$ of (\ref{shift3}) becomes 
\begin{align}	\label{shift4}
\Delta \nu 
	&= \left( \frac{1}{4 \pi \langle T \rangle} \right) 
	\sum_{\overline{\mu}, \overline{\nu}} \int_{-1}^1  \frac{ 
	f_{\overline{\mu} \, \overline{\nu}}^{\alpha \beta} (\theta) \, 
	 \phi_{\overline{\mu} \, \overline{\nu}}^{\alpha \beta} (\theta)  }{ 1 + \left[\phi_{\overline{\mu} \, \overline{\nu}}^{\alpha \beta}(\theta)\right]^2 } \, d\cos(\theta)
\end{align}
in terms of averaged molecular phase shifts 
\begin{align}
\phi_{\overline{\mu} \, \overline{\nu}}^{\alpha \beta}(\theta) = \langle [ \omega_{\overline{\mu} \, \overline{\nu}}(\theta) - \omega_{\alpha \beta} ] \tau \rangle   
\end{align} 
that generalize $m_{\mu\nu}\phi$ in previous work.\cite{mcguyer:thesis} 
The remaining angle brackets here, and subsequently, denote a rovibrational expectation value assuming a single rovibrational state. 

Unfortunately, this approximation limits the comparison of fitted values of parameters with those inferred from relaxation measurements. 
For example, the distribution of states could produce both a zero shift (\ref{shift4}) and significant damping, because the numerator is linear while the corresponding numerator for damping is nonlinear. 
Likewise, even without this approximation, the appropriate distributions may conceivably differ, since relaxation need not be coherent. 
As a result, the fitted values for $\langle T \rangle$ and $\langle N \rangle$ for the shift (\ref{shift4}) are expected to be similar, but not necessarily equal to those from relaxation measurements such as shown in Table~\ref{tab2}. 

\subsection{Interaction energies and fit parameters} 

Last, what remains is to evaluate the molecular interaction energies and choose their fit parameters. 
The results agree to leading order in $1/N$ with calculations from a bound-atom approach.  
For fitting, it is convenient to use pressure-independent and isotope-independent parameters. 
Noting that $\tau p$ is very nearly pressure independent, let us introduce a molecular phase-shift parameter 
\begin{align}	\label{PhiuvabThetaP}
\left(\phi_{\mu\nu}^{\alpha \beta} (\theta) \, p \right) &= (\phi p)_\text{hfs} 
	+ (\phi_{\mu\nu} p)_\text{sr} + \left(\phi_{\mu\nu}^{\alpha \beta}(\theta) \, p\right)_\text{Z} \nonumber \\ 
	&+ (\phi_{\mu\nu} p)_\text{dh} 
	+ (\phi_{\mu\nu} p)_\text{qh} 
\end{align} 
with pressure-independent contributions from each interaction to be determined below. 
For convenience, the bar notation is remove from the indices $\mu$ and $\nu$ here and subsequently. 
Isotope-independent fit parameters will be introduced for each contribution. 
The top line has the contributions from the three interactions previously considered, including the Zeeman (Z) interaction in (\ref{H0}). 
As shown below, only the Zeeman contribution depends on $\alpha$, $\beta$, and $\theta$. 
The bottom line has the contributions from the two interactions not previously considered.

\subsubsection{Hyperfine-shift interaction} 

The perturbation to the Fermi-contact interaction is modeled by the hyperfine-shift interaction (hfs), 
\begin{align}
V_\text{hfs} = \delta A(R) \, {\bf I} \cdot {\bf S}, 
\end{align}
where $\delta A(R)$ is a potential that depends on internuclear distance $R$. 
This interaction is diagonal in G and $m_G$, and has the same form as the Fermi-contact interaction in (\ref{H0}). 
It is the dominant spin interaction responsible for the linear pressure shift.\cite{ishikawa:2023,ishikawa:2022,camparo:2007jcp,oreto:2004} 
The potential $\delta A(R)$ is generally expected to change sign at least once at small $R$ for our molecules of interest.\cite{mcguyer:thesis,mcguyer:dA,tarakanova:2020,tscherbul:2009,camparo:2007jcp,oreto:2004,walker:1997,rao:1970,ray:1975,kleppner:1974,kleppner:1976,dA:KHe} 
As a result, the sign of the fitting parameters for binary and molecular shifts need not agree.\cite{mcguyer:2011pra} 

The molecular energies follow Eqs.~(10.49) and (11.80) in Brown and Carrington\cite{brown:book} and are the same as those in the atomic case, 
\begin{align} 
E_{| F N G g \rangle}^\text{hfs} = \langle \delta A(R) \rangle \frac{1}{4} \left\{ (F - b)[b] + (F - a)[a] \right\}, 
\end{align} 
where brackets denote a rovibrational expectation value. 

The molecular phase-shift parameter 
\begin{align}	\label{PhiPhfs}
(\phi p)_\text{hfs} &= 
	 \frac{\left\langle \delta A(R) \, \tau \right\rangle p \, [I] }{ 2 \hbar } 
	 = 2 \pi \nu_{00} (\psi p)_\text{hfs} 
\end{align}
is independent of $\mu$ and $\nu$, and equal to $(\phi p)$ in previous work.\cite{gong:2008,mcguyer:2011pra,mcguyer:thesis} 
For fitting data, the parameter 
\begin{align}	\label{PsiPhfs}
(\psi p)_\text{hfs} &= 
	 \left\langle (\delta A(R)/A) \, \tau \right\rangle p 
\end{align}
(units of s Torr) is pressure independent and isotope independent, because the ratio $\delta A/A$ is isotope independent.\cite{mcguyer:thesis} 
The value of $A$ used here and in the $\nu_{00}$ of (\ref{v00}) includes all perturbations other than the pressure shift. 
That is, the values $\nu_{00}$ are zero-pressure intercepts, though in practice, $\nu_{00}$ may be replaced by the ideal values in Table~\ref{tab1} with little impact.

\subsubsection{Spin-rotation interaction} 
The electronic spin-rotation interaction (sr) is 
\begin{align}	\label{Vsr}
V_\text{sr} = \gamma(R) \, {\bf S} \cdot {\bf N},  
\end{align} 
where $\gamma (R)$ is a potential that depends on internuclear distance.\cite{happer:book,walker:1997,wu:1985,cooke:1977,freeman:1976} 
While the literature may convey a general impression that $\langle \gamma(R) \rangle$ should be positive, experiment and theory confirm it is negative for LiAr.\cite{bruhl:2001,meshkov:2005} 
This interaction can have multiple physical origins,\cite{brown:book} but the most significant origin here is expected to be the spin-orbit interaction in the noble-gas atom.\cite{wu:1985} 
This interaction is diagonal in $G$ and $m_G$. 

The molecular energies follow Eqs.~(10.48) and (11.85) in Brown and Carrington,\cite{brown:book} 
\begin{align} 
E_{| F N G g \rangle}^\text{sr} 
	=& 
	\frac{(-1)^{F-a} \langle \gamma(R) \rangle}{[I]} \bigg( \overline{m} \langle N \rangle 
	+\frac{ \overline{m}(\overline{m} + 1) - F(F+1) }{2} \bigg) 
\end{align} 
where $\overline{m} = G - N$. 
The first term matches previous work.\cite{mcguyer:thesis} 
The second term is a correction of relative order $1/\langle N\rangle$ and resembles the dipolar and quadrupolar energies below. 

The molecular phase-shift parameters are 
\begin{align}	\label{PhiPsr}
(\phi_{\mu\nu} p)_\text{sr} &= 
	\frac{(\psi p)_\text{sr}}{[I]} \bigg( \mu + \nu  \nonumber \\ 
	&+ \frac{2 \mu (\mu + 1) + 2 \nu (\nu + 1) - 4 I (I+1) - 1}{4 \langle N \rangle } \bigg) 
\end{align}
using a pressure- and isotope-independent parameter 
\begin{align}	\label{PsiPsr}
(\psi p)_\text{sr} &= 
	 \left\langle \gamma(R) N \tau \right\rangle p / \hbar 
\end{align}
(units of rad Torr) 
that equals $r_1 \phi p [I]$ in previous work.\cite{mcguyer:2011pra, mcguyer:thesis}  

Table~\ref{tab2} provides estimates of the magnitude of $(\psi p)_\text{sr}$ from relaxation measurements. 
The sign of $(\psi p)_\text{sr}$ has an effect on $\Delta \nu$ only through the small $1/N$ correction term on the second line of (\ref{PhiPsr}). 
This correction term is not present in previous bound-atom models\cite{mcguyer:thesis,mcguyer:2011pra,camparo:2022,camparo:TOR} 
and, surprisingly, contributes to the linear shift in addition to the nonlinear shift, as shown below. 
This term depends on $(\psi p)_\text{sr}/\langle N \rangle$, so is independent of $N$ in the single-state approximation.

\subsubsection{Dipolar-hyperfine interaction} 

The electron-nuclear dipolar (or anisotropic hyperfine-shift) interaction (dh) can be expressed in many forms.\cite{brown:book} For a strong coupling of ${\bf I}$ and ${\bf S}$, a convenient form is 
\begin{align} \label{dh1} 
V_\text{dh} &= \sqrt{6} g_S \mu_B g_I \mu_N \left( \frac{\mu_0}{4 \pi} \right) T^2({\bf C}) \cdot T^2 ({\bf S}, {\bf I}), 
\end{align} 
following the notation of Eq.~(1.56) in Brown and Carrington,\cite{brown:book} where $\mu_0$ is the vacuum permeability and $T^2({\bf C})$ is a tensor involving the alkali-metal valence electron position about its nucleus. 
This interaction is diagonal in $G$ and $m_G$. 
For reference, using Eq.~(5.116) in Ref.~\onlinecite{brown:book}, an effective Hamiltonian for this interaction is  
$V_\text{dh} \approx t_0(R) \, {\bf I} \cdot (3 \hat{R} \hat{R} - \mathbb{1} ) \cdot {\bf S},$ 
where the axial magnetic-dipole hyperfine coefficient 
$t_0(R) = g_S \mu_B g_I \mu_N \left( \frac{\mu_0}{4 \pi} \right) \langle \eta, \Lambda | T^2_{0}({\bf C}) | \eta, \Lambda \rangle$ 
measures the spherical asymmetry of the Rb valence electronic wavefunction, $\hat{R}$ is the internuclear axis unit vector, and $\mathbb{1}$ is the unity dyadic tensor. 
The coefficient $t_0$ is related to the Frosh and Foley parameter $c = t_0/3$.\cite{brown:book} 

The molecular energies for (\ref{dh1}) follow Eqs.~(1.60), (8.513--8.515), (10.50), and (11.81--11.84) in Brown and Carrington.\cite{brown:book} 
The calculation includes a Wigner 9j symbol that is available in Table 10.3 of Ref.~\onlinecite{varshalovich}. 
These energies are 
\begin{align} 
&E_{| F N G g \rangle}^\text{dh} 
	=  
(-1)^{F-b} \frac{\langle t_0(R) \rangle}{2  [I]} \nonumber \\ 
	&\times \left\langle \frac{3X_{F, \overline{m}}(X_{F, \overline{m}}-1)-4N(N+1)F(F+1)}{(2N-1)(2N+3)}\right\rangle 
\end{align} 
for the quantities 
\begin{align} 
X_{F, \overline{m}} 
	&= F(F+1)-\overline{m} (\overline{m}+[N])  \label{XFm} 
\end{align} 
and $\overline{m} = G - N$, assuming $I \geq 1/2$. 
The brackets apply to the remaining dependence on $N$. 

The molecular phase-shift parameters are  
\begin{align}	\label{PhiPdh}
(\phi_{\mu\nu} p)_\text{dh} 
	&= 
	\frac{g_I (\psi p)_\text{dh}}{2 [I] } \left\langle \frac{2 N(N+1)[I]^2 - 3Y_{\mu,\nu}^+}{(2N-1)(2N+3)} \right\rangle 
\end{align} 
for the quantity 
\begin{align} 	\label{Yuv}
Y_{\mu,\nu}^\pm 	&= X_{a,\mu}(X_{a,\mu}-1) \pm X_{b,\nu}(X_{b,\nu}-1)
\end{align}
using a pressure- and isotope-independent parameter 
\begin{align}	
(\psi p)_\text{dh} &= 
	 {\left\langle t_0(R) \, \tau p \right\rangle  }/({ g_I \hbar })  
\end{align} 
(units of rad Torr). 
For reference, the leading-order term in a $1/\langle N \rangle$ expansion is 
\begin{align}	
&(\phi_{\mu\nu} p)_\text{dh} \approx 
	g_I 
	(\psi p)_\text{dh}  
	\left( \frac{[I]}{4} - \frac{3 (\mu^2 + \nu^2)}{2[I]} \right). 
\end{align}
The first part resembles the hyperfine-shift interaction and the second part modifies the shape of the nonlinear shift.

\subsubsection{Quadrupolar-hyperfine interaction} 

The nuclear electric quadrupole interaction (qh) can be expressed in many forms.\cite{brown:book} 
For a strong coupling of ${\bf I}$ and ${\bf S}$, a convenient form is 
\begin{align} \label{qh1} 
V_\text{qh} &= -e \, T^2({\bf Q}) \cdot T^2 (\nabla {\bf E}), 
\end{align}
following the notation of Eq.~(1.28) in Brown and Carrington.\cite{brown:book} 
Here, $Q$ is the quadrupole moment of the alkali-metal nucleus, with values given in Table~\ref{tab1}. 
This interaction is diagonal in $G$ and $m_G$. 
For reference, using Eqs.~(5.116), (7.158), and (7.192) in Ref.~\onlinecite{brown:book}, an effective Hamiltonian for this interaction is 
$V_\text{dh} \approx \frac{ e q_0(R) Q }{ 4I(2I-1) } \, {\bf I} \cdot (3 \hat{R}\hat{R} - \mathbb{1} ) \cdot {\bf I},$ 
where $q_0(R)$ is a standard measure of the electric-field gradient along the internuclear axis. 

The molecular energies for (\ref{qh1}) follow a derivation similar to Eq.~(9.93) and Appendix 8.4 in Brown and Carrington,\cite{brown:book} using Eqs.~(5.173), (9.13-14), and (7.159) in that reference. 
These energies are 
\begin{align} 
&E_{| F N G g \rangle}^\text{qh} 
	=  
	\frac{ e \langle q_0(R) Q \rangle }{4 I (2I-1)} \left( \frac{(-1)^{F-a}}{[I]} - \frac{1}{2} \right) \nonumber \\ 
	&\times 
	\left\langle \frac{3X_{F, \overline{m}}(X_{F, \overline{m}}-1)-4N(N+1)F(F+1)}{(2N-1)(2N+3)}\right\rangle
\end{align} 
for $\overline{m} = G - N$ and $X_{F, \overline{m}}$ of (\ref{XFm}), assuming $I \geq 1$. 
The brackets apply to the remaining dependence on $N$. 

The molecular phase-shift parameters are 
\begin{align}	\label{PhiPqh}
&(\phi_{\mu\nu} p)_\text{qh} 
	=   
	\frac{3 (\psi p)_\text{qh} Q}{8 I (2I-1) [I]}
	\left\langle
	\frac{ 2 Y_{\mu,\nu}^+ - [I] Y_{\mu,\nu}^-}{(2N-1)(2N+3)}
	\right\rangle 
\end{align}
for the quantities $Y_{\mu,\nu}^\pm$ of (\ref{Yuv}), 
using a pressure- and isotope-independent parameter  
\begin{align}	
(\psi p)_\text{qh} &= 
	 {e \left\langle q_0(R) \, \tau p \right\rangle}/{ \hbar } 
\end{align}
(units of rad Torr/Barn). 
For reference, the leading-order term in a $1/\langle N \rangle$ expansion is  
\begin{align}	
&(\phi_{\mu\nu} p)_\text{qh} \approx  
	\frac{3 (\psi p)_\text{qh} Q}{8 I (2I-1) [I]}  \left[ \nu^2 (2I+3) - \mu^2 (2I-1)  \right]. 
\end{align} 
This term modifies the shape of the nonlinear shift. 

\subsubsection{Zeeman interaction} 

The Zeeman interaction (Z) in $H_0$ has both nuclear and electronic spin contributions. 
For simplicity, as in previous work,\cite{gong:2008,mcguyer:2011pra,mcguyer:thesis} let us ignore the smaller nuclear contribution. 
Let us also ignore the slight variation in the g-factor of alkali-metal--noble-gas van der Waals molecules.\cite{freeman:1976gfactor} 
Then the Zeeman interaction is very nearly 
\begin{align}
H_\text{Z} = - \bm{\mu} \cdot {\bf B} 
	\approx g_S \mu_B \, {\bf B} \cdot {\bf S}. 
\end{align} 
This interaction sets the atomic and molecular quantization axes to be along the applied field ${\bf B}$ in the lab frame, so is diagonal in $m_G$. 
However, it is not diagonal in $G$, and for sufficiently large fields $B$, will induce mixing between low-field eigenstates with different values of $G$. 

The molecular energies follow Eqs.~(11.91--92) in Brown and Carrington.\cite{brown:book} 
For the electronic spin only, the energies are 
\begin{align} 
E_{| F N G g \rangle}^\text{Z} &= 
	(-1)^{F-a} g_S \mu_B B \left( \frac{2 g}{[G]} \right) \nonumber \\
	&\times \left(\frac{[G]}{2}\right) \frac{G(G+1)+F(F+1)-N(N+1)}{2G(G+1)}, 
\end{align} 
where the factor of $2g/[G]$ was not reduced to leave the connection with (\ref{thetaGg}) clear. 
The molecular phase-shift parameters are 
\begin{align} 	\label{PhiPz}
(\phi_{\mu\nu}^{\alpha \beta} p)_\text{Z}  
	&\approx \frac{g_S \mu_B B \langle \tau p \rangle }{[I] \hbar} \bigg[ -\alpha - \beta + \cos(\theta) \bigg( \mu + \nu \nonumber \\
	&+ \frac{ 4 I (I+1) + 1 - 2 \mu^2 - 2 \nu^2 }{4 \langle N \rangle} \bigg) \bigg], 
\end{align} 
using (\ref{sharedangle}) and (\ref{thetaGg}) and 
including only the first and second terms in a $1/N$ expansion. 
The first term is equivalent to $-(r_1 \phi p) (\alpha + \beta)$ in the low-field limit of previous work.\cite{mcguyer:thesis} 
The second term is a small correction with a form that roughly resembles the dipolar and quadrupolar energies above.

\subsection{Fitting functions} 
Finally, using the above averages and approximations, we can construct fitting functions for data analysis as follows. 
We will first do this for the general case, and then simplify the functions for the particular case of the 0--0 transition with a negligible applied field $B$, which describes the Rb data well. 

Consider a data set of $\alpha$--$\beta$ transition frequencies $\left\{ \nu(p) \right\}$ of a particular alkali-metal atom measured at different pressures $p$ of a single-species buffer gas (i.e., He or Xe, but not a mixture). 
We may fit this data set using a fitting function 
\begin{align}	\label{fitf1}
f(p) = \nu_0 + s\, p + \Delta^2\nu 
\end{align} 
with a zero-pressure intercept $\nu_0$, a total pressure-shift slope $s$, 
and a nonlinear pressure-shift function $\Delta^2 \nu$. 
Explicit forms for the nonlinear shift are given below, which use atomic and molecular parameters in Tables I and II. 

In the data analysis to follow, we will simultaneously fit pairs of two such data sets, one for $^{85}$Rb and one for $^{87}$Rb, measured separately for the same single-species buffer gas. 
To do this, we will separate the total linear slope $s = s_b + s_m$ into its binary ($s_b$) and molecular ($s_m$) contributions, and use a 
shared fitting parameter $\sigma_b$ instead of the isotope-dependent parameter $s_b$. 
The expected isotopic scaling of the binary slope follows the hyperfine-shift interaction, so a convenient choice for a pressure- and isotope-independent parameter is the fractional-frequency-shift slope 
$\sigma_b = s_b / \nu_{00}$ 
(units of 1/Torr). 
The fitting function (\ref{fitf1}) then becomes 
\begin{align}	\label{fitf2}
f(p) 
	&= \nu_0 + (s_b + s_m)\, p + \Delta^2\nu \nonumber \\ 
	&= \nu_0 + (\nu_{00} \, \sigma_b + s_m)\, p + \Delta^2\nu, 
\end{align} 
with explicit forms for the molecular slope given below. 
This fitting function is then used for each Rb-isotope data set, with separate zero-pressure intercepts $\nu_0$ for each isotope. 
The remaining molecular slopes and nonlinear shifts are evaluated appropriately for each isotope using shared isotope- and pressure-independent fitting parameters. 
Note that the same fitting parameter is used for $\langle N \rangle$ in all interactions.

\subsubsection{Fit function for moderate applied fields} 
For moderate applied-field strengths $B$ such that $G$ is still a good quantum number, the nonlinear pressure shift is given by (\ref{shift4}) with molecular phase-shift parameters given in (\ref{PhiuvabThetaP}), (\ref{PhiPhfs}), (\ref{PhiPsr}), (\ref{PhiPdh}), (\ref{PhiPqh}), and (\ref{PhiPz}). 
Re-arranging to use pressure-independent parameters, the linear molecular slope becomes 
\begin{align} 	\label{slope5}
s_m  
	&= \left( \frac{1}{4 \pi \langle Tp^2 \rangle} \right)  
	\sum_{{\mu}, {\nu}} \int_{-1}^1  
	f_{{\mu} \, {\nu}}^{\alpha \beta} (\theta) 
	\left[ \phi_{{\mu} {\nu}}^{\alpha \beta} (\theta) p \right] 
	\, d\cos(\theta), 
\end{align}
and the nonlinear shift becomes 
\begin{align}	\label{shift5}
\Delta^2 \nu 
	&= \left( \frac{-1}{4 \pi \langle Tp^2 \rangle} \right)  
	\sum_{{\mu}, {\nu}} \int_{-1}^1  \frac{ 
	f_{{\mu} \, {\nu}}^{\alpha \beta} (\theta) 
	\left[\phi_{{\mu} {\nu}}^{\alpha \beta} (\theta) p \right]^3 p }{ p^2 + \left[\phi_{{\mu} {\nu}}^{\alpha \beta} (\theta) p \right]^2 } 
	\, d\cos(\theta), 
\end{align}
where the pressure-independent parameter $\langle T p^2 \rangle$ describes molecule formation.  

The molecular slope (\ref{slope5}) evaluates to a function of the field, 
\begin{align} 	\label{smB}
s_m(B) = s_m(0) - \frac{g_S \mu_B B \langle \tau p \rangle }{3 \pi [I] \hbar \langle Tp^2 \rangle} (\alpha + \beta),  
\end{align} 
for $I \in \{ 1/2, 3/2, 5/2, 7/2\}$ and potentially higher half-integral values. 
The first term is the same as the zero-field value given in the next section, and varies with the transition. 
The second term is a high-pressure correction to the linear Zeeman splitting of the alkali-metal atom, because the coherence $\langle \alpha | \rho | \beta \rangle$ samples differing Zeeman shifts as it connects through the molecule via different $\mu$--$\nu$ pathways.

\subsubsection{Fit function for negligible applied fields} 
For magnetic fields that are small enough to not influence the nonlinear shift significantly while still defining the quantization axis, we may set $B=0$ such that 
the phase-shift parameters $(\phi_{{\mu} {\nu}}^{\alpha \beta} p )$ of (\ref{PhiuvabThetaP}) 
do not depend on $\theta$. 
Then (\ref{shift5}) becomes  
\begin{align}	\label{shiftFinal}
\Delta^2 \nu &\approx 
	\Delta^2_2 \nu = 
	- \left(\frac{1}{2\pi \langle Tp^2 \rangle} \right) 
	\sum_{{\mu}, {\nu}}  \frac{W_{{\mu} {\nu}}^{\alpha \beta} \, 
	(\phi_{{\mu}  {\nu}}^{\alpha \beta} p)^3 p }{ p^2 + (\phi_{{\mu} {\nu}}^{\alpha \beta} p)^2 } 
\end{align}
with a subscript introduced for Section~\ref{Rel2PrevWork}. 
The fitting parameters within $(\phi_{{\mu} {\nu}}^{\alpha \beta} p )$ are given by (\ref{PhiuvabThetaP}), (\ref{PhiPhfs}), (\ref{PhiPsr}), (\ref{PhiPdh}), and (\ref{PhiPqh}). 
The negligible-field weight coefficients 
\begin{align}	\label{Wuvab}
W_{{\mu} {\nu}}^{\alpha \beta} 
	&= \frac{1}{2} 
	\int_{-1}^1 
	f_{{\mu} {\nu}}^{\alpha \beta} (\theta) \, d\cos \theta 
\end{align}
generalize the $W_\sigma = \sum_\mu W_{\mu, \sigma - \mu}^{00}$ in previous work.\cite{mcguyer:thesis} 
For the 0--0 transition, 
these weights are 
\begin{align} \label{W00}
W_{{\mu} {\nu}}^{0 0} 
	&= (-1)^{\mu + \nu} \sum_{k} \frac{1}{[k]} C^{k0}_{a0;a,0} C^{k0}_{a\mu; a,-\mu} C^{k0}_{b 0; b, 0} C^{k0}_{b,\nu;b,-\nu}.  
\end{align}
Table~\ref{tab3} gives explicit values for the weights of interest here. 

The corresponding molecular linear slope (\ref{slope5}) becomes 
\begin{align} 	
s_m(0) = \left( \frac{1}{2 \pi \langle T p^2\rangle} \right)  
	\sum_{\overline{\mu}, \overline{\nu}}
	W_{{\mu} {\nu}}^{\alpha \beta} \, 
	( \phi_{\overline{\mu} \, \overline{\nu}}^{\alpha \beta} p ), 
\end{align} 
which is the zero-applied-field value of (\ref{smB}), and evaluates to 
\begin{align} \label{slopeFinal}
s_m(0) =&~\frac{ 1 }{ \langle T p^2\rangle} \bigg( \nu_{00} (\psi p)_\text{hfs} 
	- \frac{(\psi p)_\text{sr} [I]}{12 \pi \langle N \rangle } \nonumber \\ 
	&- \frac{g_I (\psi p)_\text{dh} [I] [1 + 2 I(I+1)]}{10 \pi (2\langle N \rangle-1)(2\langle N \rangle+3)} \nonumber \\ 
	&+ \frac{(\psi p)_\text{qh} Q [I] [3-4 I(I+1)]}{40 \pi I(2I-1) (2\langle N \rangle-1)(2\langle N \rangle+3)} \bigg)
\end{align} 
for $I \in \{ 1/2, 3/2, 5/2, 7/2\}$ and potentially higher half-integral values. 
Note that the quadrupolar interaction is not allowed for $I=1/2$. 
The first term from the hyperfine-shift interaction shares an isotopic scaling with $s_b \propto [I] A$, so is indistinguishable from $s_b$ except for how parameters are fitted via $\Delta^2 \nu$. 
The second term from the spin-rotation interaction has a different isotopic scaling than $s_b$, so is distinguishable by comparing isotopes as discussed in Section~\ref{SRsmDiscussion}, in which case it tends to dominate the fitted value of $\langle N \rangle$.  
The remaining contributions from the dipolar and quadrupolar interactions are of second and higher order in $1/\langle N \rangle$ and also have different isotopic scalings. 

The threshold to enter the negligible-field regime depends on the choice of transition and the fit parameter values (see Supplementary Material). 
Numerically, for the 0--0 transition, this regime seems to reliably occur for $|B| \lesssim |B_1|$ for the spin-rotation fields $B_1$ in Table~\ref{tab2}, though it often extends to larger fields. 
However, the thresholds differ between $\alpha$--$\beta$ transitions, and may be lower, for example, for end-state resonances.

\begin{table}[t]
\caption{
Weights $W_{\mu\nu}^{0 0}$ of (\ref{W00}) for the 0--0 transition 
versus nuclear spin quantum number $I$. 
Common alkali-metal atoms are indicated for convenience. 
The rows correspond to $\mu \in [a, a-1, \ldots , 1-a, -a]$, and the columns to $\nu \in [b, b-1, \ldots , 1-b, -b]$. 
\label{tab3} 
}
\begin{ruledtabular}
\begin{tabular}{ l l } 
$I$ 		& $W_{\mu\nu}^{0 0}$  	\\
\midrule 
	$1/2$ & 
$\begin{pmatrix}
1 \\
1 \\ 
1 \\  
\end{pmatrix}/3$\\ 
	$3/2~\left(^{87}\text{Rb}\right)$ & 
$\begin{pmatrix}
9    &      3     &     9    \\
6    &     9      &    6    \\
5    &    11  &       5   \\ 
6    &    9       &   6    \\
9    &      3     &     9 
\end{pmatrix}/105$ \\
	$5/2~\left(^{85}\text{Rb}\right)$ & 
$\begin{pmatrix}
50   &      20    &     25   &      20   &     50   \\
30        &  45          & 15        &  45      &    30    \\
24      &  36     &    45     &    36    &     24  \\ 
23 &      29 &      61  &     29  &     23  \\
24      &  36     &    45     &    36    &     24  \\ 
30        &  45          & 15        &  45      &    30    \\
50   &      20    &     25   &      20   &     50    
\end{pmatrix}/1155$ \\
	$7/2~\left(^{133}\text{Cs}\right)$ & 
$\begin{pmatrix}
245 & 105 & 105 & 91 & 105 & 105 & 245 \\ 
140 & 210 & 84 & 133 & 84 & 210 & 140 \\ 
110 & 150 & 204 & 73 & 204 & 150 & 110 \\ 
100 & 120 & 180 & 201 & 180 & 120 & 100 \\ 
97 & 117 & 141 & 291 & 141 & 117 & 97 \\ 
100 & 120 & 180 & 201 & 180 & 120 & 100 \\ 
110 & 150 & 204 & 73 & 204 & 150 & 110 \\ 
140 & 210 & 84 & 133 & 84 & 210 & 140 \\ 
245 & 105 & 105 & 91 & 105 & 105 & 245 
\end{pmatrix}/9009$ \\ 
\end{tabular}	
\end{ruledtabular}
\end{table}

\subsubsection{Relation to fit functions in previous work} \label{Rel2PrevWork}
Previous work focused on the 0--0 transition with negligible applied fields.\cite{gong:2008,mcguyer:2011pra,mcguyer:thesis} 
We can recover the ``low-field spin-rotation'' model of Ref.~\onlinecite{mcguyer:2011pra} from (\ref{shiftFinal}) by setting $(\psi p)_\text{dh} = 0$, $(\psi p)_\text{qh} = 0$, and $\langle N \rangle \longrightarrow \infty$. 
Using the relation 
$\sum_\mu W^{00}_{\mu, \sigma - \mu} = W_\sigma$, where $\sigma = \mu + \nu$, 
this recovers the nonlinear shift 
\begin{align}	\label{LFSR}
\Delta^2_2 \rightarrow \Delta^2_1 = - \left( \frac{1}{2 \pi T} \right)  \sum_{\sigma = -2I}^{2I} \frac{W_\sigma (1 + r_1 \sigma)^3 \phi^3}{1 + (1 + r_1 \sigma)^2 \phi^2}
\end{align} 
from Refs.~\onlinecite{mcguyer:2011pra,mcguyer:thesis}, 
where $\phi = \delta A[I] \tau/(2 \hbar)$ and $r_1 = 2 \gamma N / \left(\delta A [I]^2 \right)$. 
The fitting parameters 
$(\psi p)_\text{hfs} = 2 \pi \nu_{00} (\phi p)$ 
and 
$(\psi p)_\text{sr} = (r_1 \phi p) [I]$. 
We can recover the original model of Ref.~\onlinecite{gong:2008} by additionally setting $(\psi p)_\text{sr} = 0$. 
Using the property 
$\sum_{\sigma} W_{\sigma} = 1$, 
this recovers the nonlinear shift 
\begin{align}
\Delta^2_2 \rightarrow \Delta^2_0 = - \left( \frac{1}{2 \pi T} \right) \frac{\phi^3}{1 + \phi^2}, 
\end{align} 
of Refs.~\onlinecite{gong:2008,mcguyer:thesis}. 
In both of the above cases, this also recovers the corresponding linear molecular slope $s_m = (\phi p) / [2 \pi (T p^2)]$ of all previous work.\cite{gong:2008,mcguyer:2011pra,mcguyer:thesis,camparo:2022,camparo:TOR} 

The models of Refs.~\onlinecite{camparo:TOR,camparo:2022} discussed in Section V are equivalent to (\ref{LFSR}). 
However, those works use an admixing model to modify the isotopic scaling of the fitting parameters, and additional modifications to directly or indirectly alter the functional form with pressure. 

Ref.~\onlinecite{mcguyer:thesis} models different transitions and the effects of moderate applied fields using a bound-atom approach. 
We recover very nearly the same model by setting $(\psi p)_\text{dh} = 0$, $(\psi p)_\text{qh} = 0$, and $\langle N \rangle \longrightarrow \infty$. 
However, the details and interpretation of the angular average in (\ref{shift5}) are rather different than that of Eq.~(4.98) in Ref.~\onlinecite{mcguyer:thesis}, except in the limit of small applied fields. 
The approach in Ref.~\onlinecite{mcguyer:thesis} may offer an approximate method to extend this work to strong fields where $G$ is no longer a good quantum number.

\begin{figure}[t!]
\centering
\includegraphics[width=\columnwidth]{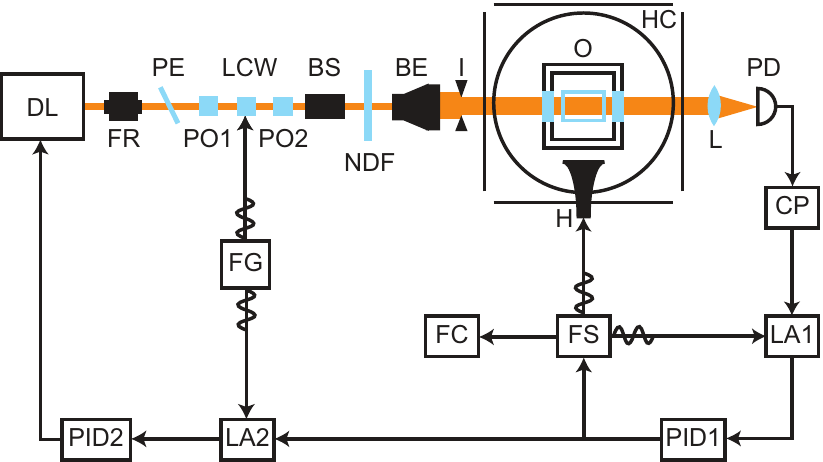}
\caption{ 	\label{fig:expt} 
Experimental setup for the measurement of nonlinear pressure shifts.
DL, diode laser; 
FR, Faraday rotator; 
PE, pellicle; 
PO, polarizer; 
LCW, liquid crystal wave plate; 
BS, beam shaper; 
NDF, neutral density filter; 
BE, beam expander; 
I, iris;
O, oven;
H, horn; 
HC, Helmholtz coils; 
L, lens; 
PD, photodetector; 
CP, current preamplifier; 
LA, lock-in amplifier; 
PID, proportional-integral-derivative (PID) controller; 
FS, frequency synthesizer;
FC, frequency counter; 
FG, function generator.} 
\end{figure}

\begin{figure*}
\centering
\includegraphics[width=2\columnwidth]{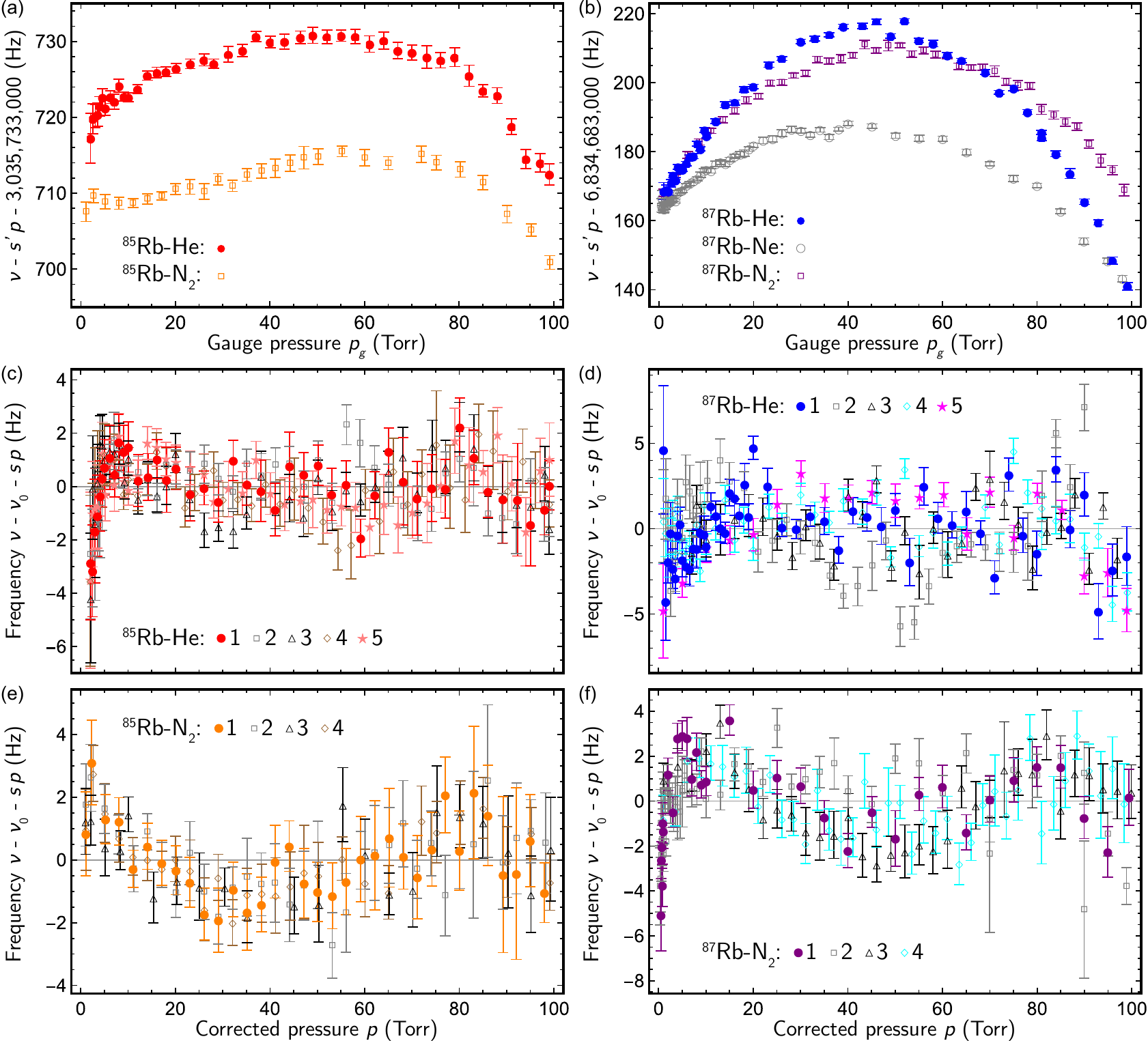}
\caption{ 	\label{fig3} 
Pressure-gauge linearization using 
the 0--0 resonant frequencies $\nu$ of $^{85}$Rb (left) and $^{87}$Rb (right) in He and N$_2$ at 40.0$^\circ$ C and $B = 1$ G. 
(a,b) Linear fitting residuals of select data sets versus raw pressure-gauge readings $p_g$ show an apparent nonlinearity. 
The zero-pressure intercepts $\nu_0$ were not removed to illustrate its typical variation and to help distinguish data sets. 
The $^{87}$Rb plot also includes Ne data.  
(c--f) The apparent nonlinearities disappear after correcting the pressure values using Eq.~(\ref{pgc}) with Table~\ref{tab-gauge}. 
Each plot shows linear fitting residuals of multiple data sets of each pair of Rb isotope and pure buffer gas after this correction. 
Tables~\ref{tab-linear} and \ref{tab-nonlinear} summarize the fitting results. 
}
\end{figure*}

\section{Experimental Methods} 
The 0--0 resonant frequencies $\nu$ of $^{85}$Rb in the pure buffer gases He, N$_2$, Ar, Kr, Xe, and $^{136}$Xe were measured using the same laser-pumped, vapor-cell clock apparatus as was used for the $^{87}$Rb data in He, Ne, N$_2$, Ar, Kr, and Xe previously reported in Ref.~\onlinecite{mcguyer:2011pra}. 
Fig.~\ref{fig:expt} sketches this apparatus, and Ref.~\onlinecite{mcguyer:thesis} describes it in additional detail. 
To adapt to $^{85}$Rb, the appartus used a larger-aperture standard-gain horn antenna and a high-power amplifier between the function generator and horn to enhance the signal as needed for some gases, especially for Xe at high pressures. 

The $^{85}$Rb and $^{87}$Rb data were measured independently using separate, isotopically enriched vapor cells. 
Multiple data sets were measured for each pairing of Rb isotope and pure buffer gas, as shown in each plot of fitting residuals below. 
All data used a vapor-cell temperature of $40.0 \pm 0.1^{\circ}$ C and an applied field of $B = 1.0$ G along the light direction by default. 
Typical temperature and field variations across the cells were estimated to be within 0.2$^\circ$ C and 5 mG. 

\subsection{Light-shift suppression} 
To mitigate systematic error from the light shift (or dynamic Stark shift), the apparatus used two nested feedback loops: an inner loop 
to lock the carrier frequency of frequency-modulated microwaves to the 0--0 transition, and an outer loop 
to lock the optical frequency of the pumping light to produce no light shift of the 0--0 transition. 
While both feedback loops are engaged, a frequency counter with 1-Hz precision, referenced to a Rb frequency standard, sampled the locked carrier to provide $\nu$. 
The sample standard deviation provided the measurement uncertainty, which increased with the 0--0 linewidth at higher pressures. 

The outer feedback loop used an intensity-modulation method described in Refs.~\onlinecite{gong:2008,mcguyer:2009apl,mcguyer:2011pra,mcguyer:thesis}. 
This method locks the optical frequency to one of two wavelengths near the 795 nm D1 transition that produce no shift, which vary with the vapor-cell conditions.\cite{levi:2016} 
To verify that the light shift was suppressed to within the measurement uncertainty, the neutral-density filter was used to temporarily adjust the laser intensity by a factor of 2--4 every few steps in pressure. 
Additionally, data sets were measured and compared for both choices of zero-shift wavelengths for each pairing of Rb isotope and buffer gas. 
Each plot of fitting residuals includes one data set that used a different zero-shift wavelength than the other sets, except for $^{85}$Rb in N$_2$.

\subsection{Pressure-gauge linearization} 
Following the $^{87}$Rb cell, as well as a $^{133}$Cs cell for a different apparatus,\cite{mcguyer:2011pra,mcguyer:thesis} 
the $^{85}$Rb cell used a dedicated capacitance manometer (MKS Instruments Baratron) to measure the pressure $p$ up to 100 Torr with a precision of roughly $\pm$0.002 Torr and an accuracy of about $0.25\%$. 
Before cell construction, the manometer was verified to agree with that of the $^{87}$Rb cell to within $0.25\%$ across the 100 Torr range. 
As with the previous cells, it was critical to mitigate nonlinearity in pressure measurement in order to precisely measure nonlinear shifts of a few Hz on top of large background, linear shifts. 

Fig.~\ref{fig3} highlights this pressure-gauge nonlinearity and its removal for both Rb cells. 
Each plot shows linear fitting residuals for buffer gases with little to no expected nonlinearity, which will be discussed in Section~\ref{HeN2}. 
The top row shows example residuals using the raw pressure $p_g$ measured by the manometer gauges. 
As shown, there is an apparent nonlinearity in each data set using the measured pressures $p_g$. 

To remove this apparent nonlinearity, the true pressures $p$ were estimated from the measured pressures $p_g$ with the empirical formula 
\begin{align} \label{pgc}
p = c_1 \left(p_g + c_2 p_g^2 + c_3 p_g^3 \right), 
\end{align} 
where the coefficients $c_2$ and $c_3$ correct quadratic and cubic nonlinearities. 
The coefficient $c_1$ was chosen to preserve the pressure-measurement accuracy by least-squared minimizing the linear bias, $\int_0^{100~\text{Torr}} (p-p_g)^2 dp_g$, giving 
$c_1 = 7 (1 + 75 c_2 + 6000 c_3)/[7 + 42000 c_2^2 + 350 c_2 (3 + 20000 c_3) + 12000 c_3 (7 + 25000 c_3)]$.  

To use Eq.~(\ref{pgc}), unique values of the coefficients were determined for each cell and applied to all of its data. 
For the $^{87}$Rb cell, as well as a $^{133}$Cs cell, these values came from fitting multiple He, Ne, and N$_2$ data sets using the empirical formula and were reported previously.\cite{mcguyer:2011pra,mcguyer:thesis} 
Likewise, for the $^{85}$Rb cell, the values came from fitting multiple He and N$_2$ data sets. 
To mitigate concerns about room temperature fluctuations, the pressure gauge of the $^{85}$Rb cell was held at $30.0^\circ$ C in a temperature-stabilized enclosure for all but the Ar data. 
The best results for the $^{85}$Rb cell 
came from including a cubic correction. 

As Fig.~\ref{fig3} shows in its bottom two rows with multiple He and N$_2$ data sets, applying the correction of Eq.~(\ref{pgc}) using the coefficients in Table~\ref{tab-gauge} to linearize the pressure gauges of each cell eliminated the apparent nonlinearities, making the remaining nonlinearities comparable to measurement uncertainties across the pressure range. 


\begin{table}[t]
\caption{%
Pressure-gauge linearization coefficients for Eq.~(\ref{pgc}). 
The Supplementary Material provide plots of these corrections. 
\label{tab-gauge} 
}
\begin{ruledtabular}
\begin{tabular}{ l c c c l } 
Cell				& $c_1$ 		& $c_2 \times 10^6$ 				& $c_3 \times 10^7$ 	& References \\  
				& 				& $\left(\text{Torr}^{-1}\right)$ 	& $\left(\text{Torr}^{-2}\right)$  \\
\midrule 
$^{85}$Rb 		& 1.00046		& \hphantom{$-$0}9.9 $\pm$ 1.9 		& $-$2.00 $\pm$ 0.12  	&  This work\\ 
$^{85}$Rb 		& 1.00128		& $-$17.1 $\pm$ 5.2					& 0					& \onlinecite{mcguyer:thesis,camparo:TOR,camparo:2022} \\ 
$^{87}$Rb 		& 1.00256		& $-$34.0 $\pm$ 2.2 					& 0 					& This, \onlinecite{mcguyer:2011pra, mcguyer:thesis,camparo:TOR,camparo:2022} 
\end{tabular}	
\end{ruledtabular}
\end{table}

\begin{table*}[t]
\caption{
Fit parameters for the linear and nonlinear pressure shifts of $^{85}$Rb and $^{87}$Rb at 40.0$^\circ$~C and $B = 1$~G. 
Fitting used the function (\ref{fitf2}) with the nonlinear shift $\Delta^2\nu$ of (\ref{shiftFinal}), 
molecular slope $s_m$ of (\ref{slopeFinal}), 
and the atomic parameters in Table~\ref{tab1}, as described in the text. 
The magnitudes of $(\psi p)_\text{sr}$ were set to the values in Table~\ref{tab2}. 
Table~\ref{tab-linear} provides equivalent binary $s_b$ and molecular $s_m$ slopes for these parameters. 
\label{tab-nonlinear} 
}
\begin{ruledtabular}
\begin{tabular}{ l c c c c c c c } 
Gas	& $\sigma_b \times 10^9$
		& $\langle Tp^2 \rangle$ 
			& $(\psi p)_\text{hfs}$ 	
				& $(\psi p)_\text{sr}$ 	
					& $(\psi p)_\text{dh}$ 	
						& $(\psi p)_\text{qh}$ 
							& $\langle N \rangle$ \\ 
 	 & (Torr$^{-1}$)
 	 	& (ms Torr$^2$) 			
 	 		& (ps Torr$^{-1}$) 		
 	 			&  (rad Torr) 		
 	 				& (rad Torr) 		
 	 					& (rad Torr Barn$^{-1}$) 
 	 						& \\ 
\midrule 
He 	& \hphantom{$-$}104.39 $\pm$ 0.26 & 0 & \hphantom{000}0 & 0\hphantom{00} & \hphantom{000}0 & 0 & $\infty$ \\  
N$_2$  
	& \hphantom{$-$0}75.80 $\pm$ 0.19 & 0 & \hphantom{000}0 & 0\hphantom{00} & \hphantom{000}0 & 0 & $\infty$ \\ 
Ar	& \hphantom{000}$-$7.4 $\pm$ 0.9\hphantom{0}  	
		& 2.1 $\pm$ 3.5 	
			& \hphantom{0}$-$0.8 $\pm$ 1.5	
				& $\pm$1.016\footnote{The sign of the spin-rotation parameter for Ar had too little of an effect on the fit parameters and results, so was not determined. Fitting used a negative sign.} 
					& \hphantom{0}$-$0.6 $\pm$ 0.4
						& 0 
							& $\infty$  \\ 
Kr 	& \hphantom{0}$-$81.65 $\pm$ 0.22 		
		& 69 $\pm$ 15 
			& \hphantom{$-$0.}16 $\pm$ 9\hphantom{.0} 
				& 9.61 
					& \hphantom{0}$-$2.8 $\pm$ 0.6 
						& 0 
							& 8.6 $\pm$ 3.1 \\ 
Xe 	&\hphantom{0}$-$169.0 $\pm$ 0.4\hphantom{0}	
		& 52.5 $\pm$ 1.6 \hphantom{~} 	 	
			& \hphantom{.}$-$289 $\pm$ 10\hphantom{.}		
				& $-$31.9 	\hphantom{000} 	
					& $-$10.2 $\pm$ 1.5 
						& 0 
							& 8.2 $\pm$ 0.6  
\end{tabular}	
\end{ruledtabular}
\end{table*}

\begin{table}[b]
\caption{
Fit parameters for the binary and molecular linear pressure shift slopes in Eq.~(\ref{fitf2}) of $^{85}$Rb and $^{87}$Rb at 40.0$^\circ$~C and $B = 1$~G. 
All values follow from Table~\ref{tab-nonlinear} using (\ref{fitf2}) and (\ref{slopeFinal}). 
\label{tab-linear} 
}
\begin{ruledtabular}
\begin{tabular}{ l c c c c c c c } 
Gas	& $s_b$ for $^{85}$Rb 				& $s_b$ for $^{87}$Rb 		& $s_m$ for $^{85}$Rb 	& $s_m$ for $^{87}$Rb \\  
	& (Hz Torr$^{-1}$) 					& (Hz Torr$^{-1}$) 			& (Hz Torr$^{-1}$) 		& (Hz Torr$^{-1}$) \\ 
\midrule 
He 	& \hphantom{$-$}316.9 $\pm$ 0.8 		& \hphantom{$-$0}713.5 $\pm$ 1.8 			& 0 					& 0 	\\ 
N$_2$	
	& \hphantom{$-$}230.1 $\pm$ 0.6		& \hphantom{$-$0}518.0 $\pm$ 1.3 			& 0 					& 0 	\\ 
Ar	& \hphantom{0}$-$22.0 $\pm$ 3.0	 	& \hphantom{00}$-$51.0 $\pm$ 6.0		
												& $-$1 $\pm$ 3 			&  \hphantom{00.}$-$3 $\pm$ 7\hphantom{.00} \\ 
Kr 	& $-$247.9 $\pm$ 0.7 				& \hphantom{0}$-$558.1 $\pm$ 1.5 		
												& $-$1.7 $\pm$ 0.9 		&  \hphantom{$-$}0.08 $\pm$ 1.03 \\ 
Xe 	& $-$513.0 $\pm$ 1.2 				& $-$1155.0 $\pm$ 3.0 		
												& $-$3.6 $\pm$ 1.3 		&  \hphantom{0.}$-$28 $\pm$ 2\hphantom{.00}							
\end{tabular}	
\end{ruledtabular}
\end{table}

\section{Results}
This section presents the analysis of the measured 0--0 frequencies of $^{85}$Rb and $^{87}$Rb using the fitting function (\ref{fitf2}) with the nonlinear shift $\Delta^2\nu$ of (\ref{shiftFinal}) and molecular slope $s_m$ of (\ref{slopeFinal}), assuming negligible applied field. 
Eqs.~(\ref{PhiuvabThetaP}), (\ref{PhiPhfs}), (\ref{PhiPsr}), (\ref{PhiPdh}), and (\ref{PhiPqh}) give the molecular phase-shift parameters in terms of pressure- and isotope-independent fitting parameters. 
Tables~\ref{tab-linear} and \ref{tab-nonlinear} summarize the fitting results. 
For comparison, the Supplementary Material summarizes the results of separate analysis of the $^{87}$Rb data in previous and of the $^{85}$Rb data in unpublished work.\cite{mcguyer:2011pra,mcguyer:thesis}

\subsection{Helium and Nitrogen} \label{HeN2}
The ability of the apparatus to accurately measure nonlinear pressure shifts was explored using He and N$_2$ gases, as well as Ne for $^{87}$Rb, which span a range of linear-shift slopes comparable to those of Ar, Kr, and Xe, as shown in Tables~\ref{tab-linear} and \ref{tab-nonlinear}. 
Fig.~\ref{fig3} shows the results for He and N$_2$. 
After correcting for nonlinearity in pressure measurement as described above, the pressure shifts of both Rb isotopes appear linear for He and N$_2$, as expected from previous work. 
That is, though He, as well as Ne, are known to form van der Waals molecules with Rb, this and previous work have been unable to detect any definitive resulting nonlinearity. 
Likewise, though N$_2$ may form molecules, and instantaneous RbN$_2$ molecules during two-body collisions are known to be important in some applications,\cite{RbN2} this and previous work have been unable to detect any definitive resulting nonlinearity. 
The linear fitting residuals in the middle and bottom rows of the figure suggest that the limit of experimental accuracy for detecting nonlinearity is roughly on the order of a few Hz across the pressure range, with an expected sensitivity to the total linear-shift slope $s$. 
The remaining suggestive nonlinearities likely represent residual systematic error. 

Fitting of the He and N$_2$ data used the fit function (\ref{fitf2}) and ignored any potential contributions from molecules, giving $f(p) = \nu_0 + s_b p$ with $s_b = \nu_{00} \sigma_b$. 
Initial analysis fitted the data for each isotope separately. 
A pairwise comparison of the fitted slopes $s_b$ between isotopes agreed with the expected isotopic scaling to within a maximum error of $\pm 0.16\%$, suggesting that the relative inaccuracy of the two Rb cell pressure gauges varied within $\pm 0.16\%$ across the data sets, which is slightly better than the direct manometer test results reported above, perhaps from linearization with (\ref{pgc}). 

Final values for the linear pressure shifts in Tables~\ref{tab-linear} and \ref{tab-nonlinear} came from joint fitting of pairs of $^{85}$Rb and $^{87}$Rb data sets. 
Multiple pairs were fitted for each gas, as enumerated in Fig.~\ref{fig3}. 
To reduce the influence of the $\pm 0.16\%$ relative inaccuracy of pressure measurement for each Rb isotope, this inaccuracy was fitted directly with a parameter $\epsilon$ by differentially rescaling the pressure values as 
$p \longrightarrow p \sqrt{1+\epsilon}$ for one isotope and $p \longrightarrow p / \sqrt{1+\epsilon}$ for the other (for He and N$_2$ data only). 
The reported values for $\sigma_b$ and $s_b$ then came from averaging results. 
The reported uncertainties are the root sum of squares (RSS) of the following: 
the fitted slope uncertainties of each pair (which includes measurement uncertainties), 
the standard error of the mean (SEM) of the fitted slopes (to capture the scatter between pairs), 
the common-mode influence of the expected $\pm0.25\%$ pressure-measurement inaccuracy, 
and the propagation of the uncertainties of the gauge linearization coefficients in Table \ref{tab-gauge}. 
The joint-fitting results agreed closely with those from analyzing isotopes separately. 

As reported before, there are curious, small anomalous frequency shifts at very low pressures, below about 1 Torr, with some gases other than Xe. 
In Fig.~\ref{fig3}, these are apparent for $^{85}$Rb in He and $^{87}$Rb in N$_2$. 
These shifts seem to be a systematic effect from poor signal-to-noise ratios, though more investigation is required to confirm.

\begin{figure}[b]
\centering
\includegraphics[width=\columnwidth]{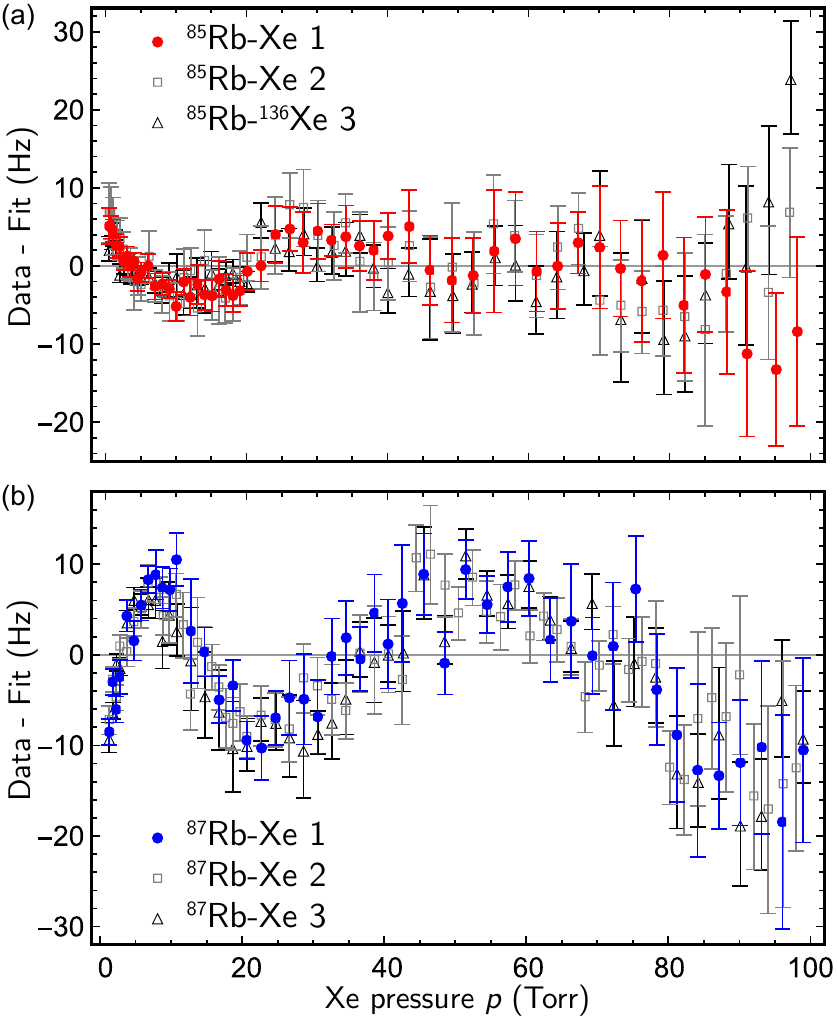}
\caption{ 	\label{fig4} 
Xenon: 
Measured 0--0 resonant frequencies $\nu$ of $^{85}$Rb and $^{87}$Rb in Xe at 40.0$^\circ$ C and $B = 1$ G. 
Fig.~\ref{fig1} shows fitted nonlinear shifts $\Delta^2\nu$ for an example pair of $^{85}$Rb and $^{87}$Rb data sets. 
(a,b) Fitting residuals for three pairs of data sets giving the parameters in Tables \ref{tab-nonlinear} and \ref{tab-linear}. 
As the top plot shows, no difference was observed between $^{85}$Rb with natural Xe and spinless $^{136}$Xe. 
} 
\end{figure}

\subsection{Xenon} 
The data for Rb in Xe is challenging to analyze because significant nonlinearity remains even at 100 Torr for both Rb isotopes, making it difficult to separate the linear and nonlinear shifts. 
Even with this limitation, the data for $^{85}$Rb in Xe challenges the models of previous work, because the shape of its nonlinearity is incompatible with that allowed by previous theory, which describes all other cases well, including $^{87}$Rb in Xe. 
Fig.~\ref{fig1} shows this visually with a comparison of the final fitted nonlinearities for both Rb isotopes reported below. 
Additionally, Ref.~\onlinecite{mcguyer:thesis} demonstrated this by fitting with previous models (see Supplementary Material). 

To probe this discrepancy, additional $^{85}$Rb data was collected with $^{136}$Xe and at $B = 0.25$ G. 
To the limit of experimental accuracy, the nonlinearities of $^{85}$Rb in natural Xe or in spinless $^{136}$Xe were the same. 
This suggests that the nuclear spin of the noble gas does not contribute significantly to nonlinear pressure shifts. 
Additionally, the relative change in mass between Xe and $^{136}$Xe is comparable to that between $^{85}$Rb and $^{87}$Rb, suggesting that this difference, and any potential impact to rovibrational distributions, is not significant. 
Likewise, no difference was observed between $B = 1$ and $0.25$ G, suggesting the Zeeman interaction is not responsible for this discrepancy. 

The model derived here suggests that this discrepancy comes from the neglect of additional spin interactions in the molecules, because the model successfully captures the shape of the nonlinearity of $^{85}$Rb in Xe. 
As Fig.~\ref{fig1} shows, the model succeeds because these interactions allow zero crossings and other freedom in the shape. 
However, the model can fit the nonlinearity by including either or both of the dipolar- and quadrupolar-hyperfine interactions. 
That is, these two interactions can produce similar nonlinearities, so the data is unable to uniquely separate their effects without additional constraints.  
Therefore, direct analysis of the $^{85}$Rb in Xe data alone gives highly non-unique results (i.e., no optimal set of fitted parameters), without further knowledge from theory. 

To test the model and make the results in Tables~\ref{tab-nonlinear} and \ref{tab-linear} as unique as possible, the fitting was constrained as follows: 
First, the values came from jointly fitting pairs of $^{85}$Rb and $^{87}$Rb data sets with shared parameters. 
Fig.~\ref{fig4} shows the fitting residuals for the three pairs used, which included one data set for $^{85}$Rb in $^{136}$Xe. 
Second, this fitting included not only the nonlinear shift, but also the dominant linear shift. 
As discussed in Section~\ref{SRsmDiscussion}, for Xe this constrained the sign of the spin-rotation parameter and required a finite value of $\langle N \rangle$. 
Third, as in previous work, the magnitude of the spin-rotation parameter was fixed to its value inferred from relaxation measurements in Table~\ref{tab2}, because the nonlinearity does not precisely determine its value. 
Fourth, only one additional spin-interaction was included, the dipolar interaction. 
Exploring the dipolar and quadrupolar interactions separately, the dipolar interaction was better able to simultaneously fit the data for both Rb isotopes. 
Including both interactions provided little benefit, so the quadrupolar interaction was neglected in the final results. 

All together, the linear and nonlinear shifts of both Rb isotopes in Xe were fitted with only the five free parameters shown in Table~\ref{tab-nonlinear}, as well as two zero-pressure intercepts $\nu_0$. 
Similar to the He and N$_2$ data, the reported uncertainties for each parameter are the RSS of the following: 
the fitted uncertainties for each pair (which includes measurement uncertainty); 
the SEM of the fitted slopes (to capture scatter between pairs); 
the differential-mode influence of $\pm0.16\%$ error between gauges (to capture inaccuracy between cell pressure gauges); 
the common-mode influence of the expected $\pm0.25\%$ pressure-measurement inaccuracy; 
and the propagation of the uncertainties of the gauge linearization coefficients in Table~\ref{tab-gauge}. 
Together, the last three sources of uncertainty strongly suggest that the results are robust against systematic errors within and between the two pressure gauges. 
Unfortunately, the zero-pressure intercepts $\nu_0$ varied at the level of a few tens of Hz between data sets, as the top row of Fig.~\ref{fig3} shows, so could not be fixed to known values to improve the fitting. 

The results in Table~\ref{tab-nonlinear} provide a minimal description of the experimental data. 
Figs.~\ref{fig1} and \ref{fig4} show that the fitting captured the data well. 
The remaining residuals are likely due to limitations in the apparatus and the model, such as neglecting rovibrational distributions. 
Separate fitting of each Rb isotope only provided modest improvements in capturing the nonlinear shapes. 
Nevertheless, without further supporting knowledge from theory, the fitted parameter results should only be considered empirical values.

\begin{figure}[t!]
\centering
\includegraphics[width=\columnwidth]{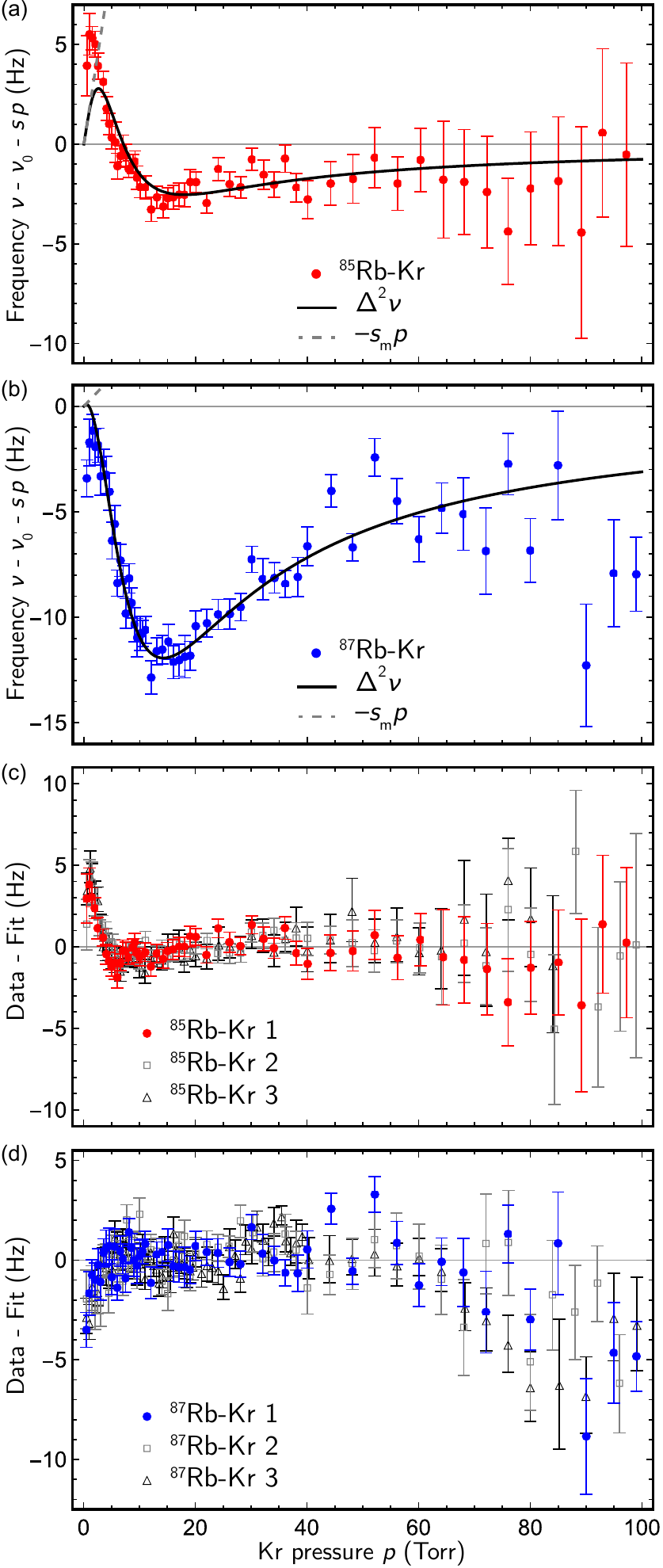}
\caption{ 	\label{fig5} 
Krypton: 
Measured 0--0 resonant frequencies $\nu$ of $^{85}$Rb and $^{87}$Rb in Kr at 40.0$^\circ$ C and $B = 1$ G. 
(a,b) Fitted nonlinear shifts $\Delta^2 \nu$ for an example pair of $^{85}$Rb and $^{87}$Rb data sets. 
(c,d) Fitting residuals for three pairs of data sets giving the parameters in Tables \ref{tab-nonlinear} and \ref{tab-linear}. 
} 
\end{figure}

\subsection{Krypton} 
Unlike for Xe, the data for Kr does appear to capture the majority of the nonlinearities below 100 Torr. 
Additionally, previous models do successfully capture the observed nonlinearity for $^{85}$Rb in Kr, as reported in Ref.~\onlinecite{mcguyer:thesis} (see Supplementary Material). 
The opportunity for Kr is then to see if the model derived here can simultaneously fit the data for both Rb isotopes, which the previous models could not do. 

Using the same approach and constraints as for Xe worked well for Kr. 
Tables \ref{tab-nonlinear} and \ref{tab-linear} report the results, which provide a minimal description of the experimental data. 
Of the nonlinear gases, Kr was the most sensitive to pressure measurement errors, because of it relatively large linear shifts and small nonlinearities. 
However, as with Xe, the fitting strongly suggest that the results are robust against errors within and between the two pressure gauges, given the uncertainties reported. 
Also, as with Xe, no difference was observed between $^{85}$Rb data at $B=1$ and $0.25$ G, supporting the use of the zero-field approximation during fitting. 

Fig.~\ref{fig5} shows that the model captures the data satisfactorily. 
However, separate fitting of the nonlinearity for $^{85}$Rb in Ref.~\onlinecite{mcguyer:thesis} (see Supplementary Material) provides some improvement, because of the relatively small size of the nonlinearity and the inability to constrain the intercepts $\nu_0$. 
Similar improvement is possible with the model derived here if the linear shifts were not fitted with shared parameters, but this would make the results highly non-unique. 
Note that, although the fitted curve for $^{87}$Rb in Kr appears similar to those of previous models, that curve has a zero crossing at low pressure, as indicated by the dashed line in Fig.~\ref{fig5}. 

\begin{figure}[t!]
\centering
\includegraphics[width=\columnwidth]{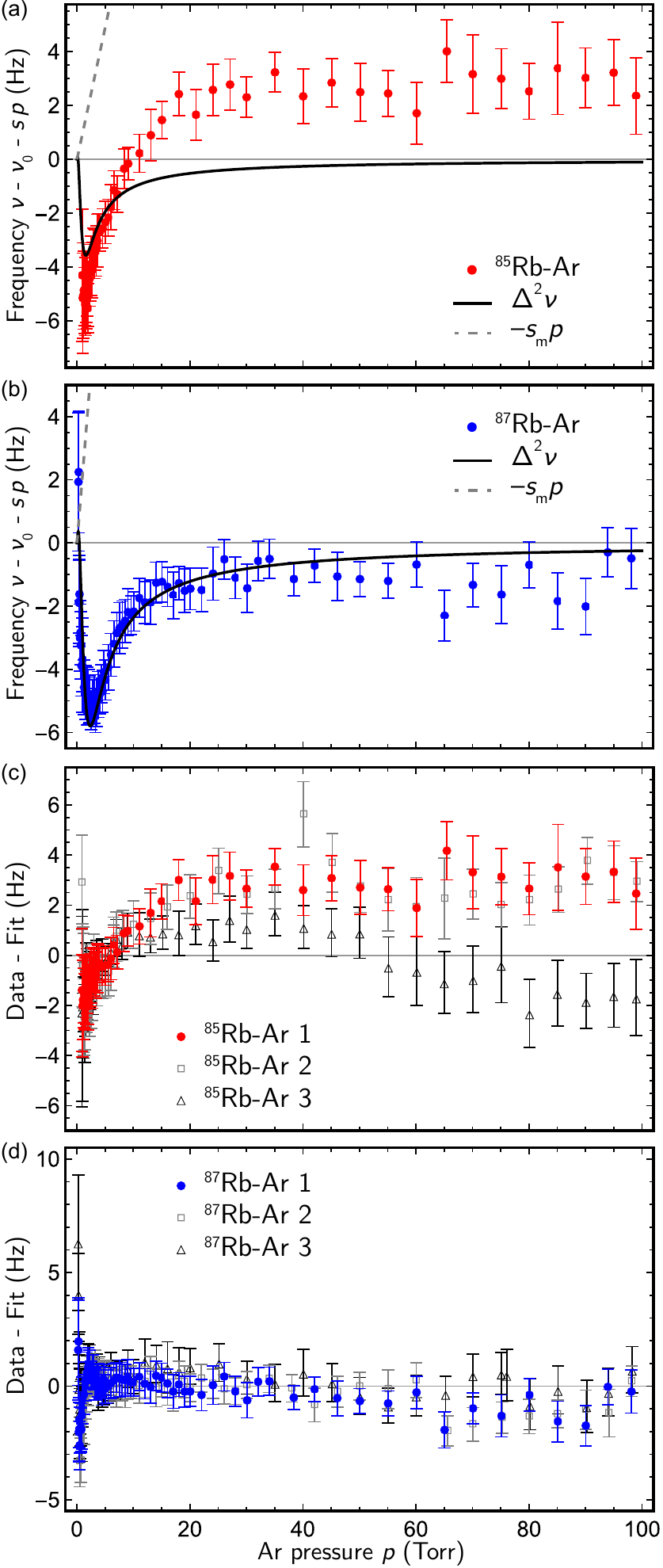}
\caption{ 	\label{fig6} 
Argon: 
Measured 0--0 resonant frequencies $\nu$ of $^{85}$Rb and $^{87}$Rb in Ar at 40.0$^\circ$ C and $B = 1$ G. 
(a,b) Fitted nonlinear shifts $\Delta^2 \nu$ for an example pair of $^{85}$Rb and $^{87}$Rb data sets. 
(c,d) Fitting residuals for three pairs of data sets giving the parameters in Tables \ref{tab-nonlinear} and \ref{tab-linear}. 
} 
\end{figure}

\subsection{Argon} 
As with Kr, the data for Ar does appear to capture the nonlinearities below 100 Torr. 
However, the the shape of the nonlinearity for $^{85}$Rb in Ar is ambiguous, because an inflection point was not captured above 1 Torr, complicating fitting. 
As a result, though previous models can fit this nonlinearity well, as reported in Ref.~\onlinecite{mcguyer:thesis}, they do so non uniquely. 
The opportunity for Ar is then the same as that for Kr, to see if the model derived here can fit the data for both Rb isotopes, which the previous models could not do. 

The same approach as that for Kr and Xe worked well for Ar. 
Tables \ref{tab-nonlinear} and \ref{tab-linear} report the results, which again provide a minimal description of the experimental data. 
However, as discussed in the next section, little to no linear molecular shift was inferred for Ar, so the sign of the spin-rotation parameter could not be determined. 
Likewise, the nonlinearities seemed insensitive to this sign, though they required a finite value for the parameter, which was again fixed to its value in Table~\ref{tab2}. 
To proceed with fitting, a negative sign was arbitrarily chosen for the spin-rotation parameter, and the parameter $\langle N \rangle$ was removed by setting its value to infinity. 
Compared to Kr and Xe, Ar was the least sensitive to pressure-measurement errors, because it has the smallest linear shifts. 
This is important because the $^{85}$Rb pressure gauge was not temperature stabilized for the Ar data. 

Fig.~\ref{fig6} shows that the model captures the data satisfactorily, given the incomplete capture of the $^{85}$Rb nonlinearity. 
Though the fitted curves resemble those of previous models, they both have zero crossings at low pressure, as indicated by the dashed lines in Fig.~\ref{fig6}. 
Parameters from separate-isotope fits are provided in the Supplementary Material.

\subsection{Spin-rotation signs using linear shifts} \label{SRsmDiscussion}
The effective sign of the spin-rotation interaction (\ref{Vsr}) is not important to relaxation, so has received little to no attention for RbAr, RbKr, and RbXe molecules. 
Unlike previous models, the model derived here predicts that this sign is particularly important to the linear shifts from molecules. 
To see this, consider measurements of the total linear, limiting slopes $s$ for both Rb isotopes and the same gas. 
Using this data, the fitting function (\ref{fitf2}) with the molecular slope (\ref{slopeFinal}) gives two equations with two unknowns, $\sigma_b$ and $(\psi p)_\text{sr}$, if the second-order contributions from the dipolar and quadrupolar interactions are neglected. 
Solving gives the parameter 
\begin{align} \label{SRsign} 
(\psi p)_\text{sr} 
	&\approx 6 \pi  \langle N \rangle \left\langle T p^2\right\rangle
	\left( \frac{  s^{87} \nu_{00}^{85} - s^{85} \nu_{00}^{87} }{ 3 \nu_{00}^{87} - 2 \nu_{00}^{85} } \right), 
\end{align} 
where superscripts indicate isotope-dependent values. 
Therefore, the sign 
$\text{sgn}[(\psi p)_\text{sr}]
	\approx \text{sgn}\left( s^{87} \nu_{00}^{85} - s^{85} \nu_{00}^{87} \right)$ 
follows from the inferred sign of any observed molecular shifts within the dominant, total linear shift slopes $s$. 

The relation (\ref{SRsign}) explains why the fitting results above were sensitive to this sign. 
Before fitting, previous results were used to estimate the signs, which predicted a negative sign for Xe with good confidence, a positive sign for Kr with reasonable confidence, but no prediction for Ar (and likewise a null result for He and N$_2$) given the $\pm0.16\%$ relative pressure-gauge inaccuracy. 
The fitting results above support these predictions. 
In particular, no solution could be found for Xe or for Kr with the other choice of sign, if both the linear and nonlinear shifts were fitted with shared parameters. 
However, the positive sign for RbKr disagrees with the theory prediction of Ref.~\onlinecite{wu:1985}. 



\section{Discussion} 

The results above show that the semi-empirical model derived in Section II provides a minimalist summary of all the data. 
For pure Ar, Kr, and Xe, which each showed a clear nonlinearity, the model successfully fit both the linear and nonlinear shifts of each Rb isotope using shared parameters. 
Attempts to use the models of previous work, excluding the modifications of Camparo\cite{camparo:TOR,camparo:2022} discussed below, failed to do this, even if the linear shifts were allowed to have separate parameters. 

The success of the new model results from two changes in its derivation, which otherwise closely resembles that of all previous models. 
The first change was to include the next two additional interactions expected to contribute, the dipolar- and quadrupolar hyperfine interactions. 
As discussed above, this was motivated in particular by the lack of any observed difference between natural Xe and spinless $^{136}$Xe, which suggested that additional spin interactions in the molecules were the source of previous discrepancy. 
Though either of the two new interactions could fit the previously incompatible shape of $^{85}$Rb in Xe, overall the dipolar interaction provided the best fit to the data of all gases. 
Additionally, the values required for the fitting parameter of the dipolar interaction were comparable to previous estimates, while those of the quadrupolar interaction were larger than expected (see Supplementary Material). 

The second change was to treat rotation quantum mechanically instead of semi-classically. 
This allowed the model to successfully fit the large linear shifts of Kr and Xe with shared parameters, because it predicted a spin-rotation contribution to the linear shift. 
While the fits are still successful without this, their results are less unique, and there would appear to be a slight disagreement between the linear shifts of the Rb isotopes beyond that expected from pressure-gauge errors. 

These changes allowed the model to fit the linear and nonlinear shifts of both Rb isotopes with only five free parameters for Kr and Xe, and four free parameter for Ar, in addition to two zero-pressure intercepts. 
The fits capture the nonlinearities well, especially for Xe, though separate-isotope fits empirically summarize those for Ar and Kr slightly better (see Supplementary Material). 
As in previous work, the single-state approximation seems adequate, though the small shifts for Ar raise concerns about competing contributions. 
The magnitudes of the spin-rotation interaction from relaxation measurements still work well for these shifts. 
However, the fitted formation rates remain only comparable to those from relaxation measurements, as expected from Section~\ref{singlestateapproximation}. 

Together, the model and results make several testable predictions for future exploration. 
The first is the set of fitting results in Table~\ref{tab-nonlinear}, which provides quantitative values to test, indicates an important contribution from the dipolar-hyperfine interaction in these molecules, and suggest a negligible contribution from the quadrupolar interaction. 
Interestingly, the fitted values of $\langle N \rangle$ for RbXe and RbKr are much lower than those in Table~\ref{tab2} from relaxation measurements. 
Further investigation could explore if low values of $N$ dominate the shift, or arise from the single-rovibrational-state approximation, or are artifacts from fitting the linear shifts using the values of $|\psi p|_\text{sr}$ from Table~\ref{tab2}. 
The second is the correction to the spin-rotation interaction energies, their contribution to the linear molecular shift, and the resulting inferred signs for this interaction for Xe and for Kr, the last of which curiously disagrees with previous theory. 
The third is the variation of the nonlinear shape with hyperfine transition, which differs from that in previous work. 
This variation seems largest between end resonances and the 0--0 transition, and should be observable with Xe (see Supplementary Material). 
The fourth is the variation of the shifts with applied magnetic field, which again differs from that in previous work. 
For low fields, the 0--0 transition shows little change until roughly 10 G for Ar and a few tens of G for Kr and Xe, which supports fitting with the negligible-field model. 
Other transitions, in particular, end resonances, show significant changes to the nonlinearity by roughly 10 G, and should be observable with all nonlinear gases (see Supplementary Material). 
Generally, the effect is to increase the size of the nonlinearity, though it can reduce it initially. 
In particular, the new model predicts that the molecules alter the linear Zeeman shift of some transitions through the linear molecular shift $s_m(B)$ of (\ref{smB}), which might be as large as roughly 36 Hz/(Torr G) for Ar. 
The fifth is the assumption that the nuclear spin of the noble gas and its spin-polarization do not matter, continued from previous work and supported here by comparing unpolarized natural Xe with spinless $^{136}$Xe.

Future theoretical and experimental work could test these predictions. 
The experimental approach used here could be readily improved to better constrain the zero-pressure interecepts, to reach lower and higher pressures, 
to probe other transitions and their field dependence, or, more generally, to probe other choices of alkali-metal and buffer-gas atoms, beyond the cases of Rb and Cs explored to date. 
As reported before,\cite{mcguyer:2011pra,mcguyer:thesis} there are curious anomalous frequency shifts at very low pressures, below about 1 Torr, with gases other than Xe. While they seem to be a systematic effect from poor signal-to-noise ratios, it is possible they could be from the nonlinearities, for example, in Kr in Fig,~\ref{fig5}. 
The apparatus remains unable to detect nonlinearities from RbHe or RbNe molecules, which are expected in the cells, or potential RbN$_2$ molecules.


Finally, two recent works by Camparo provide an alternate analysis of the $^{85}$Rb and $^{87}$Rb data in Xe.\cite{camparo:TOR,camparo:2022} 
Both works use the previous model of Refs.~\onlinecite{mcguyer:2011pra,mcguyer:thesis}, but they modify the isotopic scaling of its hyperfine-shift parameter with an admixing approach. 
This approach approximates the molecular-spin eigenstates by including hyperfine perturbations from excited Rb states, and its derivation assumes a different hyperfine propensity rule than that of Section~\ref{hyperfinepropensityrule}. 
However, the fitting functions have the same form with pressure as the previous model. 
Therefore, to accommodate the incompatible shape of the $^{85}$Rb nonlinearity in Xe, each work proposes an intriguing approach to modify the functional form. 

The first work\cite{camparo:TOR} directly modifies the form by considering a higher-order pressure dependence in the molecular lifetime $\tau$, such that the breakup-rate quantity $1/(\tau p)$ is no longer pressure independent, but instead varies between low- and high-pressure limits across the experimental range. 
However, while slight modification of the chemical dynamics is expected, for example, from corrections to the ideal gas law, the size of the modification needed to fit the data is roughly a factor of 2 for $^{85}$Rb, so seems implausible. 
That is, the collisions that form or break the molecules are expected to be isolated events with little knowledge of the surrounding pressure, within the 100 Torr measurement range. 

The second work\cite{camparo:2022} indirectly modifies the form by adding quadratic and cubic terms to the fitting function to remove a suspected error known as the ``position'' shift.  
This error comes from a variation of the 0--0 frequency with position inside the cell, which always exists at some level because of light shifts, temperature gradients, magnetic-field gradients, wall shifts, and other effects.\cite{moreno:2019,camparo:2018,camparo:1983,risley:1980} 
This variation and the way the apparatus samples it to select a single frequency likely vary with pressure, so this error may imitate a pressure shift. 
(Its related variation with microwave power is sometimes called a power shift.) 
To fit the data, this second work removed suspected positional shifts up to several kHz, with resulting nonlinearities on the order of 1 kHz (see Fig.~S3 in the Supplementary Material), similar to or larger than the inferred nonlinear shifts from molecules. 
The rather large size of these positional-shift nonlinearities seem implausible, because if they were present, then similar errors should have been observed with the other buffer gases. 
Instead, the data for He and N$_2$ shown in Fig.~\ref{fig3} (as well as Ne for $^{87}$Rb in Refs.~\onlinecite{mcguyer:2011pra,mcguyer:thesis}) suggest any positional-shift nonlinearity is on the order of several Hz. 
Likewise, the analysis of linear shifts for all gases and the observed size of the nonlinearities for Ar and Kr in Figs.~\ref{fig5} and \ref{fig6} do not support such a large effect. 
Care was taken to reduce potential sources for the positional shift in the apparatus. 
In addition to light-shift suppression described above, estimates of typical temperature and field gradients given above do not support such a large effect. 
However, wall shifts from collisions with the uncoated glass cell are indeed possible though not expected to support such a large effect either.\cite{risley:1980} 

Last, both works focused exclusively on the Xe data in the data set.\cite{vdwdata} 
Therefore, it remains to be seen whether either approach to modify the form extends to both Ar and Kr and to a detailed analysis of the linear shifts of all gases, including He and N$_2$. 


\section{Conclusion} 
In summary, measurements observed nonlinear pressure shifts for $^{85}$Rb in pure Ar, Kr, and Xe buffer gas, but not for He and N$_2$. 
These shifts revealed discrepancies with their previous modeling, in particular for $^{85}$Rb in Xe, developed with $^{87}$Rb and $^{133}$Cs data. 
Improved modeling resolved the majority of these discrepancies, and successfully fit linear and nonlinear shift data for both $^{85}$Rb and $^{87}$Rb in these gases with shared isotope-independent parameters. 
The results demonstrate the importance of the dipolar-hyperfine interaction to nonlinear pressure shifts and therefore to widely used atomic frequency standards and related atomic devices. 
Further precision measurement of such nonlinear shifts would improve the understanding of van der Waals molecules and their spin interactions.

\section{Supplementary Material} 
The Supplementary Material provides additional detail on averaging over the direction of rotation, 
estimates of the dipolar- and quadrupolar-hyperfine interaction parameters, 
a plot of the pressure-gauge linearization corrections, 
a plot of the positional shifts proposed by Camparo,\cite{camparo:2022} 
additional fitting parameters for figures above and from previous work, 
and reproductions of figures showing results from previous work.

\begin{acknowledgments} 
I am grateful to 
William Happer for suggesting this work and for many helpful discussions, 
to Michael J. Souza for making the cells, 
to Fei Gong and Yuan-Yu Jau for contributions to the original apparatus, 
to Thad G.~Walker for helpful discussions about the spin-rotation interaction and RbN$_2$ molecules, 
and to James C.~Camparo for stimulating discussions and analyses. 
I am thankful to Ben A.~Olsen, James C.~Camparo, and Yuan-Yu Jau for reading the manuscript. 
Experiments were performed at Princeton University before the author joined Amazon, with support from the Air Force Office of Scientific Research, Department of Defense, and Department of Energy. 
\end{acknowledgments} 

\section*{Data Availability Statement}
The data that support the findings of this study are openly available in the Harvard Dataverse at http://doi.org/10.7910/DVN/ZA2Z7Q (cited as Ref.~\onlinecite{vdwdata}).

\pagebreak 
~ 
\pagebreak
\setcounter{section}{0}
\renewcommand{\thesection}{S-\Roman{section}}
\begin{center}
\textbf{\large Supplementary Material of ``Isotope study of the nonlinear pressure shifts of $^{85}$Rb and $^{87}$Rb hyperfine resonances in Ar, Kr, and Xe buffer gases''}
\end{center}
\setcounter{equation}{0}
\setcounter{figure}{0}
\setcounter{table}{0}
\makeatletter
\renewcommand{\theequation}{S\arabic{equation}}
\renewcommand{\thefigure}{S\arabic{figure}}
\renewcommand{\thetable}{S\arabic{table}}

\section{Averaging over the direction of rotation} 

The rotational wave function $|\psi_N\rangle$ may be decomposed in terms of orthonormal Hund's case (b) rotational spin basis functions $|N \, n \rangle$ with $\Lambda = 0$ as 
$| \psi_N\rangle = \sum_n B_n |N \, n \rangle$, with $n = m_N$.\cite{brown:book} 
Each sticking collision then corresponds to a choice of coefficients $B_n$. 
To average over the direction of rotation, we need to determine the statistical weights $\langle B_n B_m^* \rangle$ to use with (\ref{shift2}). 

To proceed, consider using an operator ${\bf R}(\theta,\phi,\psi)$ with Euler angles $(\theta,\phi,\psi)$ to rotate an arbitrary $|\psi_N\rangle$ to a new direction. 
After this rotation, the coefficients are 
$B_n(\theta,\phi,\psi) = \sum_m B_m(0,0,0) D^{(N)}_{n, m}(\theta,\phi,\psi)$, where the Wigner ``big'' $D$-function $D^{(N)}_{n, m}(\theta,\phi,\psi) = \langle N \, n | {\bf R}(\theta,\phi,\psi) | N \, m \rangle$. 
Averaging uniformly over all Euler-angle values then gives 
$\langle B_n B_m^* \rangle = \delta_{n, m}/[N]$ from $D$-function orthogonality. 

Therefore, the substitution (\ref{subPsiN}) in Sec.~\ref{AverageOverDirection} averages over the direction of rotation.

\section{Estimates for the dipolar- and quadrupolar-hyperfine interactions} 
Ref.~\onlinecite{walker:1997} provides estimates for the strength of the dipolar- and quadrupolar-hyperfine interactions for $^{85}$Rb with $^{131}$Xe. 
We can use this to estimate parameter values as follows: 

For the dipolar interaction, noting that the coefficient $t_0(R) = B_a(R)$ in Eq.~(31) of Ref.~\onlinecite{walker:1997}, 
Fig.~14 of Ref.~\onlinecite{walker:1997} estimates $|t_0(R)| \leq |\gamma(R)|$ for $R \in [2, 4.5]$ \AA, which is just before the potential well in $V(R)$.\cite{medvedev:2018} 
Approximating equality across $R$, this gives a rough estimate of 
$|(\psi p)_\text{dh}| \leq |(\psi p)_\text{sr}|/(\langle N \rangle g_I) \approx 0.5$ rad Torr for Rb in Xe using values in Table II. 
The values in Table V are reasonably comparable to this rough estimate. 

For the quadrupolar interaction, noting that the coefficient $C_a(R) = e q_0 (R) Q / \{ 4I (2I-1)\}$ in Eq.~(31) of Ref.~\onlinecite{walker:1997}, 
Fig.~14 of Ref.~\onlinecite{walker:1997} estimates $|C_a(R)| \leq |\gamma(R)/20|$ for $R \in [2, 3.5]$ \AA, which again is just before the potential well in $V(R)$.\cite{medvedev:2018} 
Approximating equality across $R$, this gives a rough estimate of 
$|(\psi p)_\text{qh}| \leq |(\psi p)_\text{sr}|4 I(2I-1)/(20 \langle N \rangle Q) \approx 0.5$ rad Torr/Barn for Rb in Xe using values in Table II. 
Attempts to fit the nonlinear shapes of both isotopes of Rb in Xe seemed to require values about 1000 times larger than this rough estimate.

\section{Pressure-gauge linearization corrections} 
\begin{figure}[t]
\centering
\includegraphics[width=\columnwidth]{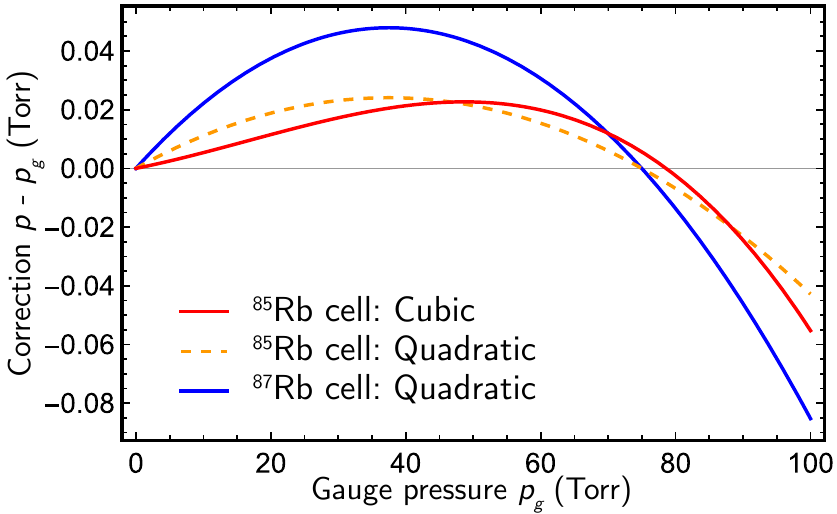}
\caption{ 	\label{figS1} 
Pressure-gauge linearization: 
The three curves show the size of the correction from the empirical formula (\ref{pgc}) across the full manometer ranges for the three sets of coefficients in Table~\ref{tab-gauge}.} 
\end{figure}

Fig.~\ref{figS1} shows the size of the corrections used to linearize the pressure gauges of both vapor cells. 

\section{Positional shift in Camparo} 
\begin{figure}[h]
\centering
\includegraphics[width=\columnwidth]{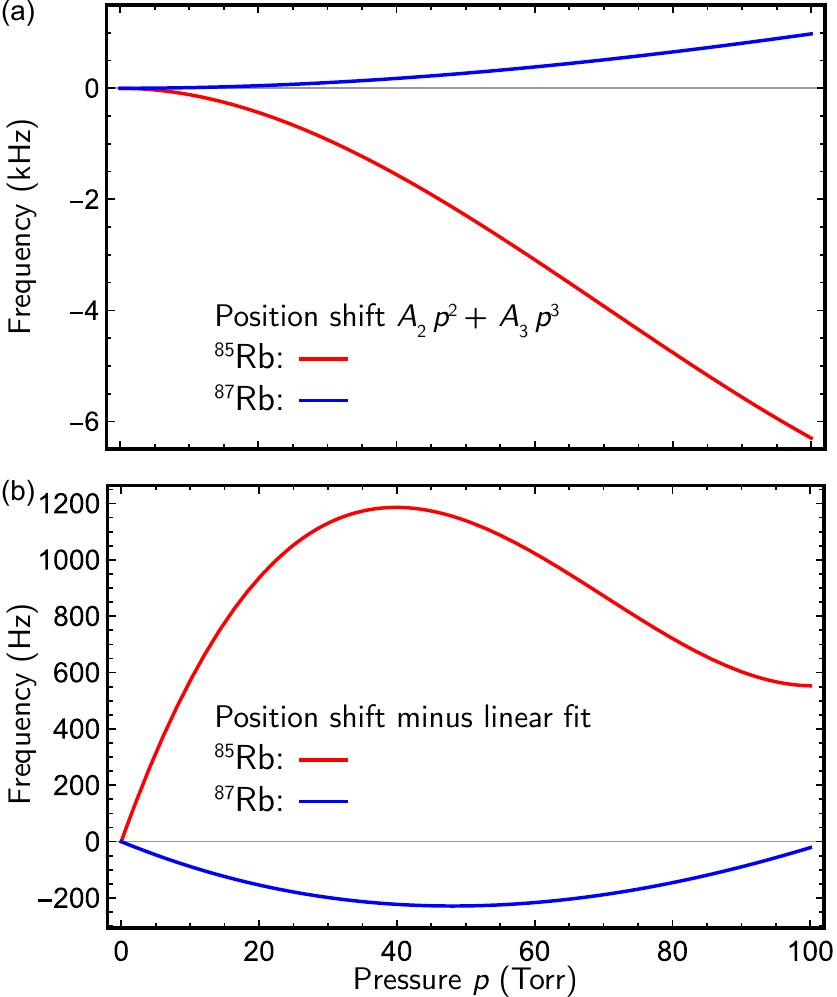}
\caption{ 	\label{figS2} 
Positional shifts in Ref.~\onlinecite{camparo:2022}: 
(a) The curves show the positional shifts proposed for the Xe data. 
The form for the curves and the reported values for its $A_2$ and $A_3$ coefficients come from Eq.~(68) and Table II in Ref.~\onlinecite{camparo:2022}. 
(b) Example nonlinearity of the proposed positional shifts inferred by linear fitting, with the zero-pressure intercepts set to zero. 
} 
\end{figure}

Fig.~\ref{figS2} shows the proposed positional shifts in the Xe data from the analysis of Ref.~\onlinecite{camparo:2022} discussed in Section V. 

\section{Predictions for variation with transition and applied field} 
\begin{figure}[t]
\centering
\includegraphics[width=\columnwidth]{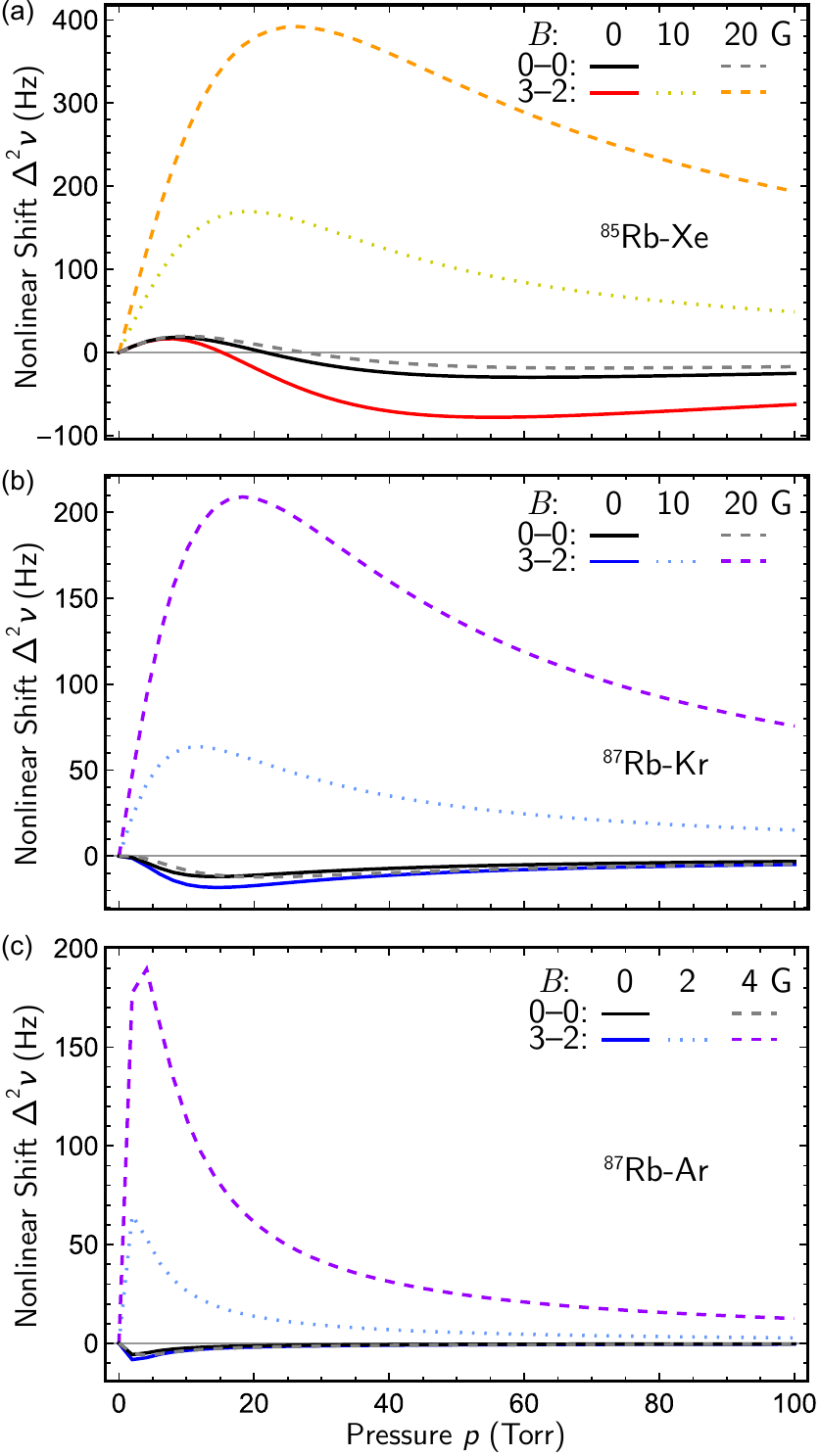}
\caption{ 	\label{figS3} 
Examples variations predicted for the nonlinearity with transition and applied field $B$: 
(a) Xenon. (b) Krypton. (c) Argon. 
The results for the Rb isotopes not shown were similar. 
At zero field, the nonlinearities varied the most between 0--0 and end resonances. 
End resonances show strong variation with applied field. 
Changing the sign of $\alpha$ and $\beta$ or $B$ often reflects the nonlinearity across the x-axis for large fields. 
} 
\end{figure}

Fig.~\ref{figS3} shows the predicted variation of the nonlinear shift with choice of $\alpha$--$\beta$ transition and applied field $B$. 
While only one Rb isotope is shown for each gas, the results are similar for the other isotopes. 
The curves come from (\ref{shift5}) using the parameters in Table~\ref{tab-nonlinear-plotted}, so that their shapes for the 0--0 transition at zero field match those shown in previous plots. 
The results provide support for fitting the 0--0 transition data at $B = 1$ G with the negligible-field approximation (\ref{shiftFinal}). 
At zero field, the variation seems most apparent when comparing end resonances with the 0--0 transition, in particular, for $^{85}$Rb in Xe. 
($^{87}$Rb in Xe shows a similar but less striking variation because of its larger zero-field nonlinearity.) 
Likewise, the variation with applied field seems largest for end resonances, with a general trend of increasing the nonlinearity, though it may decrease it initially. 
This variation does depend on the choice of transition (and field sign), with opposing end resonances appearing roughly as mirror images across $y=0$ at large fields, as expected from the property $s_m(B) - s_m(0) \propto (\alpha + \beta) B$ from (\ref{smB}). 
Of the gasses, Ar seemed most sensitive to the field, likely because the quantity 
$[s_m(B) - s_m(0)]/ [(\alpha + \beta) B]$ is roughly about 9 Hz/(Torr G) for Ar, compared to roughly 0.3 Hz/(Torr G) for the other gases. 

\section{Fit parameters for plotted curves} 
Table~\ref{tab-nonlinear-plotted} provides parameters for the fitted shifts shown in Figs.~1, 5, 6, and \ref{figS3}, which differ slightly from the values reported in Table V that resulted from analyzing multiple data sets. 
\begin{table*}[t]
\caption{
Fit parameters for the fitted curves shown in Figs.~1, 5, and 6. 
The values in Table V came from analyzing multiple data sets, so differ slightly from those given here. 
\label{tab-nonlinear-plotted} 
}
\begin{ruledtabular}
\begin{tabular}{ l c c c c c c c } 
Gas	& $\sigma_b \times 10^9$
		& $\langle Tp^2 \rangle$ 
			& $(\psi p)_\text{hfs}$ 	
				& $(\psi p)_\text{sr}$ 	
					& $(\psi p)_\text{dh}$ 	
						& $(\psi p)_\text{qh}$ 
							& $\langle N \rangle$ \\ 
 	 & (Torr$^{-1}$)
 	 	& (ms Torr$^2$) 			
 	 		& (ps Torr$^{-1}$) 		
 	 			&  (rad Torr) 		
 	 				& (rad Torr) 		
 	 					& (rad Torr Barn$^{-1}$) 
 	 						& \\ 
\midrule 
Ar	&\hphantom{10}$-$7.533 	
		& \hphantom{0}2.076 	
			& \hphantom{00}$-$0.672 
				& $-$1.016
					& \hphantom{0}$-$0.626 
						& 0 
							& $\infty$  \\ 
Kr 	&\hphantom{0}$-$81.598  		
		& 64.916 
			& \hphantom{$-$0}11.282 
				& 9.61 
					& \hphantom{0}$-$3.064 
						& 0 
							& 9.125 \\ 
Xe 	& $-$169.115
		& 54.333	 	
			& $-$293.872
				& $-$31.9 	\hphantom{000} 	
					& $-$10.470  
						& 0 
							& 8.120   
\end{tabular}	
\end{ruledtabular}
\end{table*}

\section{Fit parameters from previous work} 
\begin{figure}[t]
\centering
\includegraphics[width=\columnwidth]{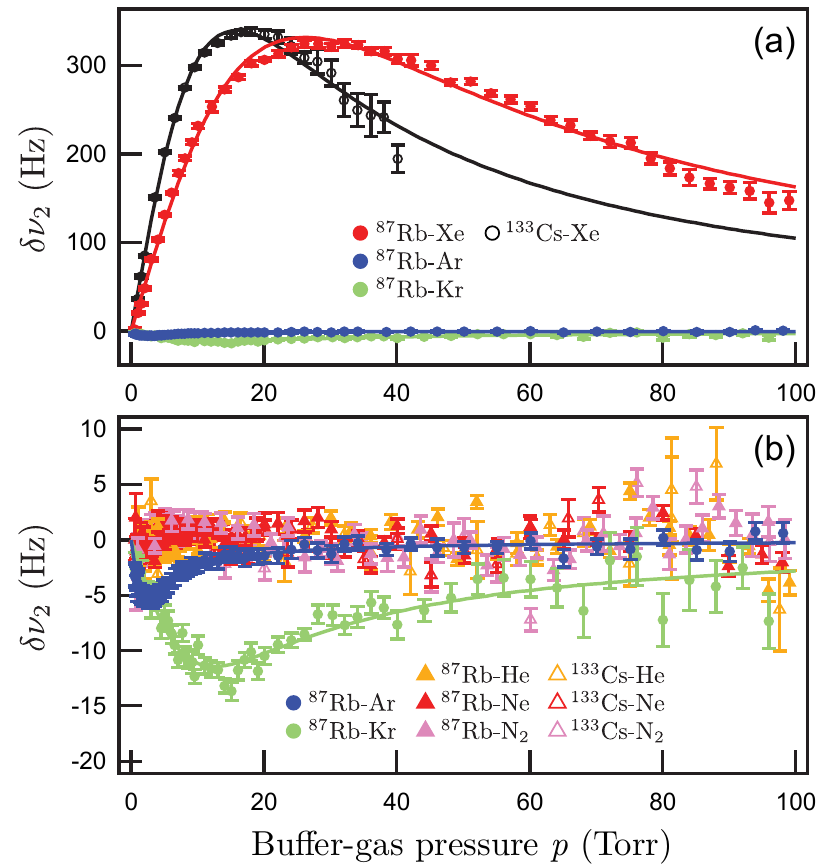}
\caption{ 	\label{figS4} 
Copy of Figure 4.6 in Ref.~\onlinecite{mcguyer:thesis} (reproduced with permission): 
Summary of the nonlinear shifts $\delta \nu_2 = \nu - \nu_0 - s p$ for $^{87}$Rb at 40.0$^\circ$ C and $B$ = 1 G and for $^{133}$Cs at 35.0$^\circ$ C and $B$ = 0.2 G as reported in Ref.~\onlinecite{mcguyer:thesis}. 
(a) $^{87}$Rb and $^{133}$Cs in Xe, with $^{87}$Rb in Ar and Kr for comparison. 
(b) $^{87}$Rb in Ar and Kr, and $^{87}$Rb and $^{133}$Cs in He, Ne, and N$_2$. 
The solid curves are the fitting results reported in Ref.~\onlinecite{mcguyer:thesis}, which Table~\ref{tab-prev-results} summarizes for $^{87}$Rb. } 
\end{figure}

\begin{figure}[t]
\centering
\includegraphics[width=\columnwidth]{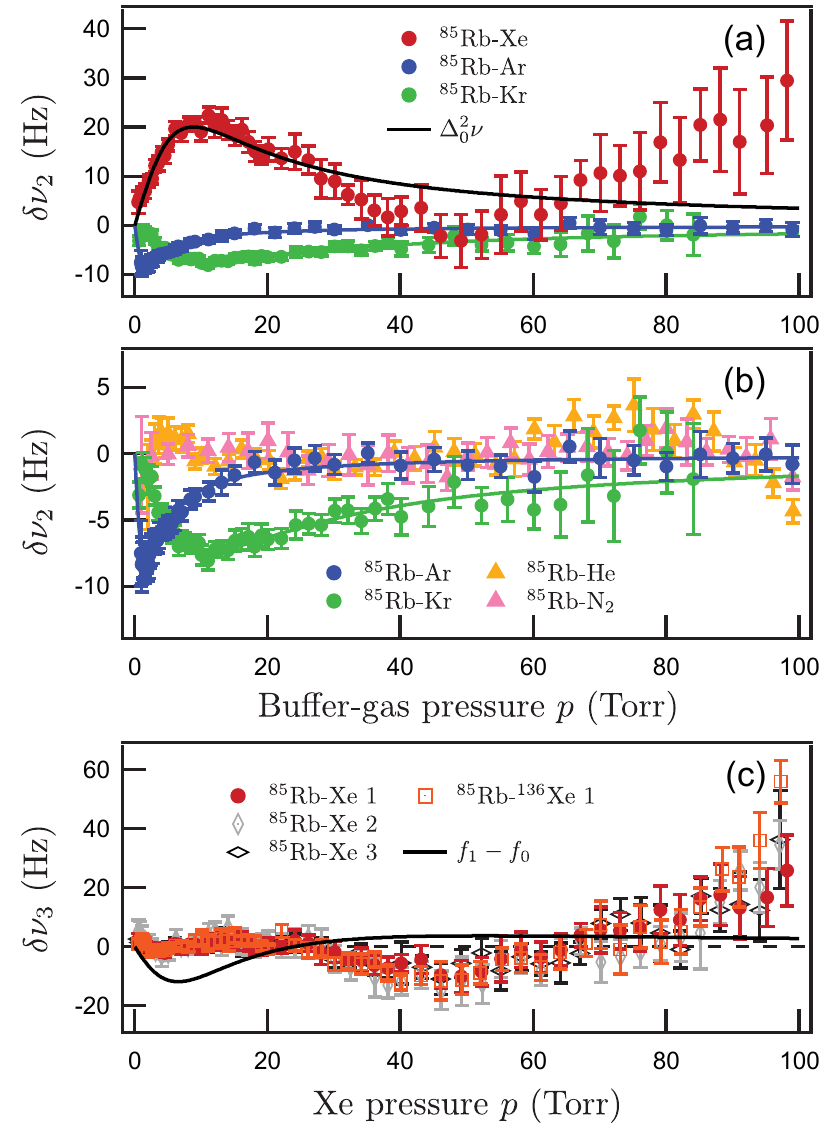}
\caption{ 	\label{figS5} 
Copy of Figure 4.7 in Ref.~\onlinecite{mcguyer:thesis} (reproduced with permission): 
Summary of the nonlinear shifts $\delta \nu_2 = \nu - \nu_0 - s p$ for $^{85}$Rb at 40.0$^\circ$ C and $B$ = 1 G as reported in Ref.~\onlinecite{mcguyer:thesis}. 
(a) $^{85}$Rb in Xe, with $^{85}$Rb in Ar and Kr for comparison. 
(b) $^{85}$Rb in Ar, Kr, He, and N$_2$. 
The solid curves in (a) and (b) are the fitting results reported in Ref.~\onlinecite{mcguyer:thesis}, which Table~\ref{tab-prev-results} summarizes. 
(c) The fit residuals $\delta \nu_3 = \delta \nu_2 - \Delta_0^2 \nu$ for $^{85}$Rb in Xe and $^{136}$Xe, for fitting without the spin-rotation interaction. The solid curve is the difference of best-fit theoretical curves, with and without the spin-rotation interaction, which Table~\ref{tab-prev-results} summarizes. 
Three separate data sets demonstrate the repeatability of the measurements in natural abundance Xe. The residuals are unchanged with isotopically enriched $^{136}$Xe. } 
\end{figure}

For reference, Table~\ref{tab-prev-results} summarizes the results of Refs.~\onlinecite{mcguyer:2011pra,mcguyer:thesis}, which analyzed the $^{85}$Rb and $^{87}$Rb data separately. 
Together, they provide an additional empirical summary of the data. 
For convenience, Figs.~\ref{figS4} and \ref{figS5} reproduce two corresponding figures in Ref.~\onlinecite{mcguyer:thesis}. 
The parameters poorly describe the nonlinearity for $^{85}$Rb in Xe, but do summarize the others well.

\begin{table*}[t]
\caption{
Summary of fit parameters from previous works that separately analyzed the $^{85}$Rb and $^{85}$Rb data.\cite{mcguyer:2011pra,mcguyer:thesis} 
The hyperfine-shift and spin-rotation parameters were converted to the isotope-independent parameters of this work. 
Uncertainties for $s$ were typically $\pm$0.25\%. 
\label{tab-prev-results} 
}
\begin{ruledtabular}
\begin{tabular}{ l l c c c c c c c } 
Gas	& Metal 
	& $s$
		& $\langle Tp^2 \rangle$ 
			& $(\psi p)_\text{hfs}$ 	
				& $(\psi p)_\text{sr}$ 	
					& $(\psi p)_\text{dh}$ 	
						& $(\psi p)_\text{qh}$ 
							& $\langle N \rangle$ \\ 
 	 & & (Hz Torr$^{-1}$)
 	 	& (ms Torr$^2$) 			
 	 		& (ps Torr$^{-1}$) 		
 	 			&  (rad Torr) 		
 	 				& (rad Torr) 		
 	 					& (rad Torr Barn$^{-1}$) 
 	 						& \\ 
\midrule 
He		& $^{85}$Rb & \hphantom{$-$0}316.8\hphantom{0} 
	& 0 & 0 & 0 & 0 & 0 & $\infty$ \\ 
He		& $^{87}$Rb & \hphantom{$-$0}714.2\hphantom{0} 
	& 0 & 0 & 0 & 0 & 0 & $\infty$\\ 
Ne		& $^{87}$Rb & \hphantom{$-$0}387.3\hphantom{0}
	& 0 & 0 & 0 & 0 & 0 & $\infty$\\ 
N$_2$	& $^{85}$Rb & \hphantom{$-$0}229.6\hphantom{0}
	& 0 & 0 & 0 & 0 & 0 & $\infty$\\ 
N$_2$	& $^{87}$Rb & \hphantom{$-$0}518.0\hphantom{0} 
	& 0 & 0 & 0 & 0 & 0 & $\infty$\\ 
Ar	& $^{85}$Rb 
	& \hphantom{00}$-$23.85 
		& 27.1 $\pm$ 4.5 
			& 88.1 $\pm$ 7.3 
				& 0
				& 0 & 0 & $\infty$ \\ 
Ar	& $^{87}$Rb 
	& \hphantom{00}$-$53.71  
		& \hphantom{0}70 $\pm$ 12 
			& 51.5 $\pm$ 3.7 
				& 0
				& 0 & 0 & $\infty$ \\ 
Kr	& $^{85}$Rb 
	& \hphantom{0}$-$249.7\hphantom{0} 
		& 1500 $\pm$ 450
			& \hphantom{0}610 $\pm$ 100
				& 0
				& 0 & 0 & $\infty$ \\ 
Kr	& $^{87}$Rb 
	& \hphantom{0}$-$558.1\hphantom{0} 
		& 1080 $\pm$ 310
			& 291 $\pm$ 44 
				& 0
				& 0 & 0 & $\infty$ \\ 
Xe	& $^{85}$Rb 
	& \hphantom{0}$-$517.7\hphantom{0}  
		& 310 $\pm$ 75
			& $-$461 $\pm$ 58\hphantom{$-$}  
				& 0  & 0 & 0 & $\infty$ \\ 
Xe	& $^{87}$Rb 
	& $-$1183.7\hphantom{0}
		& \hphantom{0}164 $\pm$ 13\hphantom{0} 
			& $-$610 $\pm$ 28\hphantom{$-$}  
				& 0  & 0 & 0 & $\infty$ \\ 
Xe	& $^{85}$Rb 
	& \hphantom{0}$-$517.9\hphantom{0} 
		& 2.2 
			& $-$0.94 
				& $-$31.9  & 0 & 0 & $\infty$ \\ 
Xe	& $^{87}$Rb 
	& $-$1184.0\hphantom{0}
		& \hphantom{0}82 $\pm$ 10 
			& $-$345 $\pm$ 35\hphantom{$-$} 
				& $-$31.9  & 0 & 0 & $\infty$ 
\end{tabular}	
\end{ruledtabular}
\end{table*}


%

\end{document}